\documentclass[preprint2,10pt,a4paper]{aastex}
\usepackage{amsmath}
\usepackage{graphicx}
\usepackage{multicol}
\usepackage{times}
\begin{document}

\title{\bf{Magnetar Oscillations II: spectral method }}

\altaffiltext{a}{Leiden University, Leiden Observatory and Lorentz Institute, P. O. Box 9513, NL-2300 RA Leiden}
\altaffiltext{b}{School of Physics, Monash University, P.O. Box 27, VIC 3800, Australia}
\author{ Maarten van Hoven \altaffilmark{ a } and Yuri Levin \altaffilmark{a,b} }
\email{vhoven@strw.leidenuniv.nl, yuri@strw.leidenuniv.nl}
\begin{abstract} \noindent
The seismological dynamics of magnetars is largely determined by a strong hydro-magnetic coupling between the solid crust and the fluid core. In this paper we set up a "spectral" computational framework in
which the magnetar's motion is decomposed into a series of basis functions which are associated with the
crust and core vibrational eigenmodes.
A general-relativistic formalism is presented for evaluation of the core Alfven modes in
the magnetic-flux coordinates, as well for eigenmode computation of a strongly magnetized crust of finite thickness. By considering coupling of the crustal modes to the continuum of Alfven modes in the core, we construct a fully relativistic dynamical model of the magnetar which allows: i) Fast and long simulations without numerical dissipation. ii) Very fine sampling of the stellar structure. 
We find that the presence of strong magnetic field in the crust results in localizing of some high-frequency crustal
elasto-magnetic modes with the radial number $n\ge 1$ to the regions of the crust where the field is nearly horizontal. While the hydro-magnetic coupling of these localized modes to the Alfven continuum in the core is reduced, their energy is drained on a time-scale of $\ll 1$ s. Therefore the puzzle of QPOs with frequencies larger than $600$ Hz still stands.

\end{abstract}

\section{Introduction}
Magnetar oscillations have been subject of extensive theoretical research since the discovery of quasi-periodic oscillations (QPOs) in the light curves of giant flares from soft gamma repeaters (SGR) (Israel et al. 2005; Strohmayer \& Watts 2005; Watts \& Strohmayer 2006; see also Barat et al. 1983). The observed oscillations are measured with high signal-to-noise ratios during time intervals of typically few minutes in the frequency range between 18 and 1800 Hz. It has been proposed by many authors that the physical origin of the QPOs are seismic vibrations of the star; an idea which opens the possibility to perform asteroseismological analysis of neutron stars, giving a unique observational window into the stellar interior. Initially it was hypothesized that the observed oscillations originate from torsional shear modes which are confined in the magnetar crust (e.g. Duncan 1998, Piro 2005; Watts \& Strohmayer 2006; Samuelsson \& Andersson 2007; Watts \& Reddy 2007; Steiner \& Watts 2009). If this hypothesis were true, then the observed QPOs would strongly constrain physical parameters in the neutron star crust. However, it was soon realized that, due to the presence of ultra strong magnetic fields ($B \sim 10^{14} - 10^{15} $ G; Kouveliotou et al., 1999) which are frozen both in the crust and the core of the star, the crustal motion is strongly coupled to the fluid core on timescales $\ll 1$ s Levin (2006, hereafter L06). Over the years several authors have studied the coupled crust-core problem (Glampedakis, Samuelsson \& Andersson 2006; Levin 2007, hereafter L07; Gruzinov 2008; Lee 2008; van Hoven \& Levin 2011, hereafter vHL11; Gabler et al. 2011a;  Colaiuda \& Kokkotas 2011; Gabler et al. 2011b). In particular L06 and L07 argued that for sufficiently simple magnetic field configurations (i.e. axisymmetric poloidal fields), the Alfven-type motions on different flux surfaces are decoupled so that the Alfven frequencies in the core feature a continuum. This result is well known from previous magnetohydrodynamic (MHD) studies, and it applies to general axisymmetric poloidal-toroidal magnetic fields (Poedts et al. 1985). It allows one to describe the problem of magnetar dynamics in terms of discrete crustal modes that couple to a continuum of Alfven modes in the core. With this approach, L07 and vHL11 demonstrated that the presence of an Alfven continuum has some important implications for magnetar oscillations: (i) Global modes of the star with frequencies that are located inside the continuum undergo strong exponential damping (this phenomenon is often called \textit{resonant absorption} in the context of MHD (Goedbloed \& Poedts 2004)). (ii) After the initial period ($<$ 1 s) of exponential decay, the system tends to settle in a steady state in which it oscillates at frequencies close to the edges of the continuum; these oscillations correspond to the so-called \textit{edge-modes}, that were first seen numerically in L07 and Gruzinov 2008, and were explained analytically in vH11. The edge-modes were further observed in the simulations of Gabler et al. (2011a) Colaiuda \& Kokkotas (2011) and Gabler et al. (2011b). \\

In the past half-decade, two distinct computational strategies have been
applied to the problem of calculating magnetar oscillations. 

(1) Several groups employed general relativistic MHD grid codes to simulate the dynamics of magnetized
neutron stars. Sotani et al. 2008; Colaiuda et al. 2009 and Cerd\'{a}-Dur\'{a}n et al. 2009 were able to reproduce continuum Alfven modes in the purely fluid stars with axisymmetric poloidal magnetic field, which provided important benchmark tests for
the ability of the codes to handle complex MHD oscillations.
Building on this, Gabler et al. (2011a), Colaiuda \& Kokkotas (2011) and Gabler et al. (2011b) included a crust in their neutron star models, and were thus able to study the coupled dynamics of the crust and the core.
(2) Our group (L07 and vHL11) and Lee (2008) decomposed the
motion of a magnetar into a set of basis functions, and studied the dynamics of the coefficients of these series expansion; we shall refer to this strategy as the "spectral method". This framework is able to handle
both the dynamical simulations and the stationary eigenmode problem; the latter reduces to solving the eigenvalue problem for a large matrix. L07 and vH11 chose the basis functions so that the crustal motion is decomposed into the normal modes of the free crust, and the core motion is decomposed into the
sum of core Alfven modes and a separate contribution of the core's "dc" displacements in reaction to
the motion of the crust. We refer the reader to Sections 3.2 of L07, 4.2 of vHL11, and 4.2 of this paper for 
technical details. This choice of basis functions casts  the dynamics of magnetars as a problem of coupled harmonic oscillators, in which the discrete modes of the crust are coupled to the Alfven modes in the core.

The computations of vH11 have been performed using Newtonian equations of motion and in the limit of a thin crust. In this paper we improve on vHL11 in two ways: 1) We adapt a realistic crust of finite thickness, threaded with a strong magnetic field. 2) We employ fully relativistic equations governing the motion of axial perturbations in the crust and the core.
Our spectral method has several practical
and conceptual advantages: (i) it is numerically inexpensive, making long simulations of the magnetar dynamics implemented on an ordinary workstation possible. (ii) It allows one to sample the stellar structure at high spatial resolution. (iii) It does not suffer from the problem of numerical viscosity that occurs in some finite difference schemes (scaling with the grid size), and
it is able to handle arbitrary axisymmetric poloidal fields, and not just those that are the
solutions of the Grad-Shafranov Equations\footnote{The approach developed by Sotani et al (2008) and used 
in Colaiuda et al. (2009, 2011)  casts the MHD equations in the core into a particularly simple 
form; see section 4.4 of Sotani et al. (2008). This transformation is possible if the poloidal field is the
solution of the Grad-Shafranov (GS) equation. There is, however, no compelling reason why the GS
equation should hold, since neutron stars feature  very strong stable stratification due to the radial gradients
in proton-to-neutron ratios (Goldreich \& Reisenegger 1992, Mastrano et al., 2011)}

The paper's plan is as follows. In section 2 we derive relativistic equations describing the magnetic forces acting on axial perturbations inside a neutron star with an axi-symmetric poloidal magnetic field. We construct a coordinate system which has one of its axes parallel to the fieldlines. The equations thus obtained will in later sections when we calculate elasto-magnetic modes of the crust, and when we calculate the Alfven continuum in the core.\\
In section 3.1 we introduce a formalism which allows us to calculate general relativistic elasto-magnetic eigenmodes of the crust by expanding the elasto-magnetic equations of motion in a set of basisfunctions. This reduces the eigenmode problem of the crust to a matrix eigenvalue problem. In sections 3.2 and 3.3 we work out the relativistic equations describing the magnetic and elastic restoring-force densities in the curved space-time of the neutron star crust. In section 3.4 we apply these equations to the formalism of section 3.1 in order to find free crustal eigenmodes and -frequencies. \\
In section 4, we find the core continuum Alfven modes in full general relativity, and we calculate their coupling to the crustal modes of section 3. The magnetar model constructed in this way, qualitatively shows the same features of the vHL11 model, i.e. above the fundamental Alfven frequency of $\sim 20$ Hz, the frequency domain is covered by the core continuum which effectively acts to damp crustal motion. For particular choices of the field configuration, the continuum may contain a number of gaps, generally well below 200 Hz. These gaps give rise to the characteristic 'edge-modes' of vHL11. Moreover, the crustal modes that reside inside gaps remain undamped. 
In the appendix we revisit the problem of crustal mode damping due to the presence of an Alfven continuum, by analytically calculating damping rates according to Fermi's golden rule. \\

\section{Relativistic equations for magnetic forces}

\textbf{\textit{Magnetic coordinates}}\\
We shall consider
strongly sub-equipartition $B\ll 10^{18}$G magnetic fields, so that the physical deformation of the
star is very small and the space-time is spherically-symmetric with respect to the star's center.
The metric can be written in the standard Schwarzschild-type coordinates $r$, $\theta$ and $\phi$.
It is natural, in analogy with the Newtonian treatments, to introduce the flux coordinate system in which one of the axes is parallel to the magnetic field lines (the precise meaning of this construction in relativity is described below). In the axisymmetric poloidal field geometry the magnetic field lines are located in planes of constant azimuthal angle $\phi$, which allows us to define the two 'magnetic' coordinates $\chi (r,\theta)$ and $\psi(r,\theta)$, such that the (covariant) vectors $\vec{e}_{\phi} = \partial / \partial \phi$ and $\vec{e}_{\chi} = \partial / \partial \chi$ are orthogonal to $\vec{e}_{\psi} = \partial / \partial \psi$. In the flux coordinate system the metric is given by
\begin{eqnarray}
ds^2 = - g_{tt} dt^2 + g_{\chi\chi} d\chi^2 + g_{\psi\psi} d\psi^2\\ 
+ 2g_{\psi\chi}d\chi d\psi + g_{\phi\phi} d\phi^2, \nonumber
\label{A1}
\end{eqnarray}
while the magnetic-field vector is given by
\begin{equation}
\vec{B}=B^{\chi}\vec{e}_{\chi}.
\end{equation}
Here $\vec{B}$ is the 4-vector whose components are given by
\begin{equation}
B^{\mu}={1\over 2}\epsilon^{\mu\nu\alpha\beta}F_{\alpha\beta}v_{\nu},
\end{equation}
and $v_{\nu}$ is the 4-velocity vector which for the stationary star is given by $v_t=g_{tt}v^t=\sqrt{-g_{tt}}$,
$v_i=0$.

Clearly, $g_{tt}$ and $g_{\phi\phi}$ are identical to the corresponding Schwarzschild metric terms,
\begin{eqnarray}
g_{tt} &=& 1 - \frac{2 m(r)}{r} \nonumber \\
g_{\phi\phi} &=& r^2 \sin^2\theta
\label{A2}
\end{eqnarray}
\\
\textbf{\textit{Maxwell's equations}}\\
The evolution of the magnetic field is described by Maxwell's equations. In curved space-time these read
\begin{eqnarray}
F_{\mu\nu ; \lambda} + F_{\lambda\mu ; \nu} + F_{\nu\lambda ; \mu} = 0
\label{A8}
\end{eqnarray}
In the ideal MHD limit, the electric field $E_{\mu} = v^{\nu} F_{\mu\nu}$ vanishes so that the only contribution to the electromagnetic tensor comes from the magnetic field:
\begin{eqnarray}
F_{\mu\nu} = - \epsilon_{\mu\nu\lambda\sigma} v^{\lambda} B^{\sigma}
\label{A9}
\end{eqnarray}
After some manipulation, the relations (\ref{A8}) and (\ref{A9}) yield the MHD equations for the magnetic field:
\begin{eqnarray}
\left( v^{\mu}B^{\nu} - v^{\nu}B^{\mu} \right)_{;\mu} = 0.
\label{A10}
\end{eqnarray}
This equation entails both magnetic induction, which describes the flux freezing that characterizes magnetic fields in the ideal MHD approximation, and Gauss' law for magnetic fields, i.e. $\left( v^{\mu}B^t - v^tB^{\mu} \right)_{;\mu} = 0$. For a static equillibrium, i.e. $v_t = \sqrt{-g_{tt}}$ and $v_i = 0$ (where the index $i$ runs over the spatial indices), Gauss' law can be expressed in the more familiar form
\begin{eqnarray}
B^i_{;i} = \frac{1}{\sqrt{g}} \left( \sqrt{g} B^i \right)_{,i}= 0
\label{A10b}
\end{eqnarray}
where $g \equiv \det{(g_{ij})}/g_{tt}$. This expression provides  basis for a convenient map between magnetic fields of Newtonian and relativistic stars. In the Newtonian case, the flux coordinates
$\chi$ and $\psi$ are functions of $r$ and $\theta$; we keep this functional form for the relativistic
versions of $\chi$ and $\psi$.
The expression in Eq (\ref{A10b}) is valid both in the curved space-time and in the flat Euclidean space (with $g_{ij}$ replaced by the Euclidean metric terms) of the Newtonian star. We can therefore use Eq (\ref{A10b}) to convert the values of the Euclidean field, $B_E$, to the correct values of the magnetic field in curved space-time, $B_S$ (the subscript $E$ stands again for \textit{Euclidean}, $S$ for \textit{Schwarzschild}): Eq. (\ref{A10b}) gives $\left( \sqrt{g_S} B_S^i \right)_{,i} = \left( \sqrt{g_E} B_E^i \right)_{,i} = 0$. We thus obtain
\begin{eqnarray}
B_S^{\chi} = \frac{\sqrt{g_E}}{\sqrt{g_S}} B_E^{\chi} = \frac{1}{\sqrt{g_{rr}}}B_E^{\chi}
\label{A10c}
\end{eqnarray}
which results in the relativistic poloidal magnetic field which is tangent to the flux
surfaces $\psi=const$ and which satisfies the Gauss' law.
(In the following we will drop the subscript $S$.) In this work, for concreteness, we take use the Newtonian configuration of the magnetic field generated by a current loop inside the neutron star and discussed in detail
in vHL11. Other Newtonian configurations are readily mapped onto the relativistic configurations using
the procedure that is specified above.\\
 
\textbf{\textit{Euler equations}}\\
The equations of motion are obtained by enforcing conservation of momentum, i.e. by projecting the conservation of energy-momentum 4-vector on the space normal to the 4-velocity $v^{\lambda}$
\begin{eqnarray}
h^{\lambda}_{~\mu} T^{\mu \nu}_{~~;\nu} = 0 
\label{A4}
\end{eqnarray}
where the projection tensor $h^{\lambda}_{~\mu}$ is given by
\begin{eqnarray}
h^{\lambda}_{~\mu} = \delta^{\lambda}_{~\mu} + v^{\lambda} v_{\mu}  
\label{A5}
\end{eqnarray}
$T^{\mu \nu}$ is the stress-energy tensor for a magnetized fluid in the ideal MHD approximation, and can be expressed as
\begin{eqnarray}
T^{\mu \nu} = \left( \rho + P + \frac{B^2}{4\pi} \right) v^{\mu} v^{\nu} + \left( P + \frac{B^2}{8\pi}\right) g^{\mu \nu} - \frac{B^{\mu} B^{\nu}}{4\pi}
\label{A6}
\end{eqnarray}
Here, $\rho$ and $P$ are the mass-density and pressure and $B^2 = B^{\mu}B_{\mu}$ is the square of the magnetic field, where $B_{\mu} = \frac{1}{2}\epsilon_{\mu\nu\lambda\sigma} u^{\nu} F^{\lambda\sigma}$ is the covariant  component of the Lorentz invariant magnetic field 4-vector ($\epsilon_{\mu\nu\lambda\sigma}$ is the four dimensional Levi-Civita symbol and $F^{\lambda\sigma}$ is the electromagnetic tensor). The equations of motion become
\begin{eqnarray}
\left( \rho + P + \frac{B^2}{4\pi} \right) v^\mu_{~;\nu}v^{\nu} =  ~~~~~~~~~~~~~~~~~~~\nonumber \\
h^{\mu\lambda} \left( P + \frac{B^2}{8\pi} \right)_{;\lambda} + h^{\mu}_{~\sigma} \left( \frac{B^{\sigma} B^{\lambda}}{4\pi} \right)_{;\lambda} 
\label{A7}
\end{eqnarray}
Here we have used the relation $v_{\nu}v^{\nu} = g_{\mu\nu}v^{\mu}v^{\nu} = -1$. Eq. (\ref{A7}) together with equation (\ref{A10}) provides a full description of (incompressible) motion of the magnetized fluid in a neutron star.\\

\textbf{\textit{Perturbation equations}}\\
We are now ready to derive equations that describe the linearized motion of a small Lagrangian fluid displacement $\zeta^{\mu}$ about the static background equillibrium of the star. The perturbed components of the velocity and the magnetic field 4-vectors, $v^{\mu}_{\rm pert}$ and $B^{\mu}_{\rm pert}$ are 
\begin{eqnarray}
v^{\mu}_{\rm pert} &=& v^{\mu} + \delta v^{\mu} = v^{\mu} + \frac{\partial \zeta^{\mu}}{\partial \tau}\nonumber\\
B^{\mu}_{\rm pert} &=& B^{\mu} + \delta B^{\mu}
\label{A11}
\end{eqnarray}
where the first terms on the right hand side denote the unperturbed equillibrium quantities, and the second terms on the right hand side denote the Eulerian perturbations associated with the displacement $\zeta^{\mu}$. In our 'magnetic' coordinates the only non-zero component of the unperturbed magnetic field is $B^{\chi} = B/\sqrt{g_{\chi\chi}}$, and because the equillibrium star is static and non-rotating the only non-zero component of the 4-velocity is $v^t = 1/\sqrt{-g_{tt}}$. Restricting ourselves to axi-symmetric torsional oscillations of the star, we introduce a small incompressible axisymmetric displacements $\zeta^{\phi}$. This implies that $v^{\mu}_{\rm{pert}~;\mu} = \delta v^{\mu}_{~;\mu} = \delta v^t_{~;t}$, and that the perturbations in pressure $\delta P$ and mass-density $\delta \rho$ vanish. Technically, a full description of the linearized motion of a neutron star would involve perturbations of the metric $g_{\mu\nu}$, requiring one to augment the above equations of motion with the perturbed Einstein equations. However, since we're considering incompressional axial oscillations only, the metric perturbations are dominated by the current dipole moment. One can show that this causes perturbations in the off-diagonal elements of the metric tensor which are of order $\delta v^2$, so that the metric perturbations can be safely ignored (the so-called Cowling approximation). Taking these considerations into account, we linearize Eq's (\ref{A7}) and (\ref{A10}) and after some work we obtain
\begin{eqnarray}
\left( \rho + P + \frac{B^2}{4\pi} \right) \frac{\partial^2 \zeta^{\phi}}{\partial t^2} = \sqrt{\frac{g_{tt}}{g_{\chi\chi}}} \frac{B}{4\pi g_{\phi\phi}}\frac{\partial}{\partial \chi} \left( g_{\phi\phi} \sqrt{-g_{tt}} \delta B^{\phi} \right)
\label{A12}
\end{eqnarray}
and
\begin{eqnarray}
\delta B^{\phi} = \frac{B}{\sqrt{g_{\chi\chi}}} \frac{\partial \zeta^{\phi}}{\partial \chi} 
\label{A12b}
\end{eqnarray}
These equations can be combined into a single one. After restoring a factor of $c^2$, we find 
\begin{eqnarray}
\left( \rho + \frac{P}{c^2} + \frac{B^2}{4\pi c^2} \right) \frac{\partial^2 \xi}{\partial t^2} = ~~~~~~~~~~~~~~~~~~~~~~~~~~~~~~~~~~~~~~~\nonumber \\ 
\sqrt{\frac{g_{tt}}{g_{\chi\chi}}} \frac{B}{4\pi c^2 \sqrt{g_{\phi\phi}}} \frac{\partial}{\partial \chi} \left[ \sqrt{\frac{g_{tt}}{g_{\chi\chi}}} g_{\phi\phi} B \frac{\partial}{\partial \chi} \left( \frac{\xi}{\sqrt{g_{\phi\phi}}} \right) \right]
\label{A13}
\end{eqnarray}
where $\xi = \sqrt{g_{\phi\phi}}\zeta^{\phi}$ is the physical displacement (in the $\phi$-direction) in unit length. This equation describes Alfven waves, traveling along magnetic field lines in the curved space-time of a magnetar. We checked that in the non-relativistic limit Eq. (\ref{A13}) reduces to the correct expression for Alfven waves in self-gravitating magnetostatic equillibria (Poedts et al., 1985). \\

\section{Modes of a magnetized crust in General Relativity}
In this section we will describe a formalism that allows us to calculate relativistic eigenmodes and -frequencies of a neutron star crust of finite thickness and realistic equation of state, threaded with an arbitrary magnetic field. By considering a crust of finite thickness, we will obtain high frequency radial harmonics that are not present in the crust model of vHL11 but which should be taken into account in view of the observed high frequency QPO's. In the past several authors carried out theo-retical analyses of torsional oscillations of neutron stars with a magnetized crust. Piro (2005), Glampedakis et al. (2006) and Steiner \& Watts (2009) considered horizontal shear waves in a plane-parallel crust threaded by a vertical magnetic field, whereas Sotani et al. (2008), Gabler et al. (2011a), Colaiuda \& Kokkotas (2011) and Gabler et al. (2011b), performed grid-based simulations of spherical, relativistic stars with dipole magnetic fields. Lee (2008) on the other hand, studied the Newtonian dynamics of spherical magnetic neutron stars, by decomposing the perturbed quantities into a set of basis functions, and following the dynamics of the expansion coefficients. Here we follow a strategy which is closely related to that of Lee (2008). In this section, we consider normal modes of the 'free' magnetized neutron star crust, i.e. in the absence of external forces. The idea in this section is to decompose the perturbed quantities into a set of orthogonal basis functions. By substituting this expansion in the equation of motion, we obtain equations for the evolution of the expansion coefficients. The solution of the crustal eigenmode problem, are in this way reduced to a matrix eigenvalue problem. The hydromagnetic coupling of the crust normal modes obtained in this section, to the core Alfven modes, will be discussed in section 4.\\

\textbf{\textit{Formalism for finding crustal eigenmodes}}\\
In a magnetized and elastic crust, the motion of a small torsional Lagrangian displacement away from equillibrium $\vec{\bar{\xi}}(\vec{x},t)$ (we use the notation from vHL11; $\bar{\xi}$ denote crustal displacements, $\xi$ denote displacements in the core), can be described in the general form
\begin{equation}
\frac{\partial^2 \vec{\bar{\xi}} }{\partial t^2} = \vec{L}_{\rm el} ( \vec{\bar{\xi}} ) + \vec{L}_{\rm mag} ( \vec{\bar{\xi}} )
\label{Eq1}
\end{equation}
where $\vec{L}_{\rm el}$ and $\vec{L}_{\rm mag}$ are the accelerations due to the elastic and magnetic forces acting on the displacement field. Expressions for $\vec{L}_{\rm el}$ and $\vec{L}_{\rm mag}$ are given and discussed in the next sub-section.  Augmented with no-tangential-stress conditions $\delta T_{r\phi}=\delta T_{r\theta}=0$ on the inner- and outer boundaries, this equation describes the free oscillations of a magnetized neutron star crust. Our procedure for solving Eq. (\ref{Eq1}) is as follows:\\

First, we decompose the crustal displacement field $\vec{\bar{\xi}} (t, \vec{x})$ into a set of basis functions $\vec{\Psi}_i (\vec{x})$,
\begin{equation}
\vec{\bar{\xi}} (t, \vec{x}) = \sum_{i=1}^{\infty} a_i (t) \vec{\Psi}_i (\vec{x}).
\label{Eq2}
\end{equation}
The functions $\vec{\Psi}_i$ form an orthonormal basis for a Hilbert space with inner product
\begin{equation}
\langle \vec{\eta}~ \vert ~\vec{\zeta} \rangle = \int_{\mathcal{V}} w(\vec{x}) ~\vec{\eta}~ \cdotp \vec{\zeta}~ d^3 x
\label{Eq3}
\end{equation}
where $\vec{\eta}$ and $\vec{\zeta}$ are arbitrary functions defined in the volume $\mathcal{V}$ of the crust, and $w(\vec{x})$ is a weight function. Orthonormality of $\vec{\Psi}_i (\vec{x})$ implies that $\langle \vec{\Psi}_i~ \vert ~\vec{\Psi}_j \rangle = \delta_{ij}$, where $\delta_{ij}$ is the Kronecker delta. The coefficients $a_i$ of the expansion of Eq. (\ref{Eq2}) are then simply $a_i (t) = \langle \vec{\bar{\xi}} (t, \vec{x}) ~\vert ~ \vec{\Psi}_i (\vec{x}) \rangle$.\\

The next step is decompose the acceleration field of Eq. (\ref{Eq1}) into basis functions $\vec{\Psi}_i$ according to Eq. (\ref{Eq2}), and to calculate the matrix elements $\langle \partial^2 \vec{\bar{\xi}}/\partial t^2 ~\vert ~ \Psi_j \rangle$. This yields equations of motion for $a_i(t)$:
\begin{equation}
\ddot{a}_j = M_{ij} ~a_i,
\label{Eq4}
\end{equation}
where the double dot denotes double differentiation with respect to time, and where
\begin{equation}
M_{ij} = \left( \langle \vec{L}_{\rm el} (\vec{\Psi}_i)~\vert ~ \vec{\Psi}_j \rangle + \langle \vec{L}_{\rm mag}(\vec{\Psi}_i) ~\vert ~ \vec{\Psi}_j \rangle \right), \nonumber
\end{equation}
Clearly, a crustal eigenmode with frequency $\omega_m$ (i.e. $a_{m,i} \propto e^{i\omega_m t}$ for all $i$), is now simply an eigenvector of the matrix $M$ with eigenvalue $-\omega_m^2$ 
\begin{equation}
-\omega_m^2 a_{m,j} = M_{ij} ~a_{m,i}.
\label{Eq4b}
\end{equation}
The index $m$ is used to label the different solutions to the above equation. In practical calculations, one truncates the series of Eq. (\ref{Eq2}) at a finite index $i=N$, so that one obtains a total number of $N$ eigensolutions. The eigenvalue problem of Eq. (\ref{Eq4}) with finite ($N \times N$) matrix $M$ can be solved by means of standard linear algebra methods. Given a set of suitable basis functions, the eigenvectors and eigenvalues (or crustal eigenfrequencies) converge to the correct solutions of Eq. (\ref{Eq1}) for sufficiently large $N$ (see the discussion of section 3.5).\\

\textbf{\textit{Orthogonality relation for elasto-magnetic modes}}\\
In the limit of $N \rightarrow \infty$, the elasto-magnetic eigenfunctions are
\begin{equation}
\vec{\bar{\xi}}_m (\vec{x}) = \sum_i a_{m,i} \vec{\Psi}_i (\vec{x}),
\label{Eq4c}
\end{equation}
where we omitted the time-dependent part $e^{i\omega_m t}$, on both sides. The eigenfunctions $\vec{\bar{\xi}}_m$ will form a new basis for a Hilbert space of crustal displacements. We can introduce an inner product $\langle ... \vert ... \rangle_{\rm me}$ in which this basis is orthogonal as follows: Consider a deformation $\vec{\bar{\xi}}(\vec{x}, t)$ of the crust, decomposed into a sum of eigenfunctions
\begin{equation}
\vec{\bar{\xi}} (\vec{x}, t) = \sum_m b_m (t) \vec{\bar{\xi}}_m (\vec{x}),
\label{Eq4d}
\end{equation}
where we incorporated the harmonic time dependence in the coefficients $b_m(t)$. Since $\vec{\bar{\xi}}_m$ are the eigenmodes of the crust, the kinetic energy of the displacement field $K (\vec{\bar{\xi}})$ must be equal to the sum of kinetic energies of the individual modes $K (b_m \vec{\bar{\xi}}_m)$
\begin{equation}
K \left( \vec{\bar{\xi}} (\vec{x}, t) \right) = \sum_m K \left( b_m (t) \vec{\bar{\xi}}_m (\vec{x}) \right).
\label{Eq4e}
\end{equation}
In the static Schwarzschild space-time of the neutron star, the conjugate time-like momentum $p_t = -E$ is a constant of geodesic motion (see e.g. Misner, Thorne \& Wheeler (1973), \S 25.2). In terms of the locally measured energy $E_{\rm L} = \sqrt{-g_{tt}} p^t$, the conserved "redshifted" energy is $E = -p_t = \sqrt{-g_{tt}} E_{\rm L}$. Similarly, the kinetic energy $K$ in terms of the locally measured kinetic energy $K_L$ is 
\begin{eqnarray}
K \left( \vec{\bar{\xi}} \right) = \sqrt{-g_{tt}} K_L \left( \vec{\bar{\xi}} \right) = \nonumber \\ 
\frac{1}{2}\int_{\mathcal{V}} \sqrt{-g_{tt}}\tilde{\rho} \left| \frac{\partial \vec{\bar{\xi}} }{\partial \tau} \right|^2 d \tilde{V} = \frac{1}{2}\int_{\mathcal{V}} \frac{\tilde{\rho}}{\sqrt{-g_{tt}}} \left| \frac{\partial \vec{\bar{\xi}} }{\partial t} \right|^2 d \tilde{V}\\
\equiv \frac{1}{2} \langle \partial \vec{\bar{\xi}} / \partial t ~\vert~ \partial \vec{\bar{\xi}} / \partial t \rangle_{\rm me} \nonumber
\label{Eq4f}
\end{eqnarray}
where $\tilde{\rho} = \left( \rho + P/c^2 + B^2/4\pi c^2 \right)$ is the mass-density in a local Lorentz frame, and $d\tilde{V} = \sqrt{g_{rr}g_{\phi\phi}g_{\theta\theta}}~dr ~d\phi~ d\theta$ is the locally measured space-like volume element. By substituting this expression for the kinetic energy into Eq. (\ref{Eq4e}), one finds that the cross-terms, $ \langle \partial \vec{\bar{\xi}}_m / \partial t ~\vert~ \partial \vec{\bar{\xi}}_k / \partial t \rangle_{\rm me} = \omega_m \omega_k \langle \vec{\bar{\xi}}_m ~\vert~ \vec{\bar{\xi}}_k  \rangle_{\rm me}$  with $m \neq k$, vanish. After normalizing the eigenfunctions $\vec{\bar{\xi}}_m$, so that $K (b_m \vec{\bar{\xi}}_m) = 1/2 \omega_m^2 b_m^2$, we obtain the orthogonality relation:
\begin{eqnarray}
\langle \vec{\bar{\xi}}_m ~\vert~  \vec{\bar{\xi}}_k \rangle_{\rm me} = \int_{\mathcal{V}} \frac{\tilde{\rho}}{\sqrt{-g_{tt}}} \vec{\bar{\xi}}_m \cdotp \vec{\bar{\xi}}_k  d \tilde{V} = \delta_{mk}.
\label{Eq4g}
\end{eqnarray}
The coefficients $b_m (t)$ are now simply obtained by taking the inner product between the displacement field $\vec{\bar{\xi}} (\vec{x},t)$ and the eigenfunctions $\vec{\bar{\xi}}_m (\vec{x})$:
\begin{eqnarray}
b_m (t) = \langle \vec{\bar{\xi}}(\vec{x},t) ~\vert~  \vec{\bar{\xi}}_m (\vec{x}) \rangle_{\rm me}.\\ \nonumber
\label{Eq4h}
\end{eqnarray}

In the next two sections we give expressions for $\vec{L}_{\rm mag}$ and $\vec{L}_{\rm el}$, and we discuss our choice of basis functions $\vec{\Psi}_i$ and the resulting boundary forces (due to the no-stress boundary conditions) at the end of section (3.2). In section (3.3) we set up a realistic model of the magnetar crust and we calculate the corresponding elasto-magnetic modes in section (3.4), where we apply the formalism described above. In the remainder of this paper, we will focus solely on axi-symmetric azimuthal displacement fields, i.e. $\vec{\bar{\xi}} = \bar{\xi} ~\hat{e}_{\phi}$ (where $\hat{e}_{\phi}$ is the unit vector in the azimuthal direction and $\bar{\xi}$ is the displacement amplitude) and $\partial \bar{\xi} / \partial \phi = 0$. \\

\subsection{Magnetic force density in the free crust}

While the equations of section 2 hold at arbitrary locations in the star, we will now consider magnetic forces acting on axi-symmetric, azimuthal perturbations $\vec{\bar{\xi}} (r, \theta) = \bar{\xi} (r, \theta) \hat{e}_{\phi}$ in the 'free' crust, i.e. a crust with no external stresses acting on it. This implies that to Eq. (\ref{A13}) we have to add boundary force terms arising from this no-external-stress condition. The tangential forces per unit area on both boundaries are given by
\begin{eqnarray}
T_{\rm mag}(r_{\rm in} + \epsilon) - T_{\rm mag}(r_{\rm in} - \epsilon) &=& T_{\rm mag}(r_{\rm in} + \epsilon)\\
T_{\rm mag}(r_{\rm out} + \epsilon) - T_{\rm mag}(r_{\rm out} - \epsilon) &=& -T_{\rm mag}(r_{\rm out} - \epsilon) \nonumber
 \label{Eq8a}
\end{eqnarray} 
where $T_{\rm mag}(r)$ is the magnetic stress at $r$, and $\epsilon$ is an infinitesimal number. Adding the boundary terms, we obtain
\begin{eqnarray}
L_{\rm mag} (\bar{\xi}) = ~~~~~~~~~~~~~~~~~~~~~~~~~~~~~~~~~~~~~~~~~~~~~~~~~~~~~~~~~~~~~~~~~ \nonumber\\
\sqrt{\frac{g_{tt}}{g_{\chi\chi}}}\frac{B}{4 \pi c^2 \tilde{\rho}\sqrt{g_{\phi\phi}}}\frac{\partial}{\partial \chi} \left[ \sqrt{\frac{g_{tt}}{g_{\chi\chi}}} g_{\phi\phi} B \frac{\partial}{\partial \chi} \left( \frac{\bar{\xi}}{\sqrt{g_{\phi\phi}}} \right)   \right] \\
 + \frac{1}{\tilde{\rho}}T_{\rm mag} \left[ \delta (r - r_0) - \delta (r - r_1) \right] \nonumber
 \label{Eq8}
\end{eqnarray} 
where  the $\delta$'s are Dirac delta functions. The magnetic stress $T_{\rm mag}$ is derived by linearizing Eq. (\ref{A6}) and retaining first order terms. One obtains
\begin{eqnarray}
T_{\rm mag} = \frac{\sqrt{g_{tt}g_{\phi\phi}}}{g_{\chi\chi}}\cos{\alpha} \frac{B^2}{4\pi} \frac{\partial}{\partial \chi} \left( \frac{\bar{\xi}}{\sqrt{g_{\phi\phi}}} \right)
\label{Eq9}
\end{eqnarray}

\subsection{Relativistic equations for elastic forces}

In the following we use relativistic equations describing the elastic force density acting on axial perturbations in the crust as derived by Schumaker \& Thorne (1983) (see also Karlovini \& Samuelsson 2007), 
and presented in a convenient form by Samuelsson \& Andersson (2007, SA) (for more details on the derivation of the following equations we refer the reader to these two papers). 
As shown in SA, the equation of motion for axial perturbations in a purely elastic crust, i.e. $\partial^2 \vec{\bar{\xi}}/ \partial t^2 = \vec{L}_{\rm el}(\vec{\bar{\xi}})$, can be solved by expanding the displacement field $\vec{\bar{\xi}}(r,\theta, \phi)$ into vector spherical harmonics $\vec{\bar{\xi}}_{\rm{H}, \textit{lm}}(\theta, \phi) \propto \vec{r} \times \vec{\nabla} Y_l^m$ (where $Y_l^m$ is a spherical harmonic of degree $l$ and order $m$) and corresponding radial- and time-dependent parts $\bar{\xi}_{\rm{R}}(r)$ and $f_T(t)$ of the displacement field. Rewriting Eq. (2) of SA now gives

\begin{eqnarray}
\vec{L}_{\rm el} \left( \vec{\bar{\xi}} \right) = ~~~~~~~~~~~~~~~~~~~~~~~~~~~~~~~~~~~~~~~~~~~~~~~~ \nonumber \\
\frac{1}{\tilde{\rho}}\left[ \frac{1}{r^3} \sqrt{\frac{g_{tt}}{g_{rr}}} \frac{d}{d r} \left( \sqrt{\frac{g_{tt}}{g_{rr}}}r^4\mu\frac{d}{d r} \left( \frac{\bar{\xi}_{\rm{R}}}{r} \right)\right) \right. \\
\left. - \mu g_{tt} \frac{(l-1)(l+2)}{r^2} \bar{\xi}_{\rm{R}} \right] ~\vec{\bar{\xi}}_{\rm{H}, \textit{lm}}~f_T \nonumber
\label{Eq10}
\end{eqnarray} 
where the metric terms $g_{tt}$ and $g_{rr}$ are the standard Schwarz-schild metric terms, and $\mu (r)$ is the (isotropic) shear modulus. The expansion of $\vec{\bar{\xi}}$ into vector spherical harmonics, leads to a particularly simple stress-free boundary condition for the radial function $\bar{\xi}_{\rm R}$:
\begin{eqnarray}
\frac{d}{dr} \left( \frac{\bar{\xi}_{\rm R}}{r} \right) = 0
\label{Eq11}
\end{eqnarray} 
which is valid on the inner- and outer boundaries, $r=r_0$ and $r=r_1$.\\

We are now ready to select our basis functions $\vec{\Psi}_i$ in order to solve Eq. (\ref{Eq1}). It is convenient to seperate $\vec{\Psi}_i$ into angular and radial parts, i.e. $\vec{\Psi}_i = \vec{\Psi}_{\rm{H}, \textit{i}}~\Psi_{\rm{R}, \textit{i}}$. Although our particular choice of basis is technically arbitrary, in view of the above discussion a natural choice for the angular part $\vec{\Psi}_{\rm{H}, \textit{i}}$ are vector spherical harmonics of order $m=0$ and $l=2,4,6...~etc.$ (we consider axi-symmetric perturbations which are anti-symmetric with respect to the equator),
\begin{eqnarray}
\vec{\Psi}_{\rm{H}, \textit{l}} (\theta) = \sqrt{\frac{4\pi}{l(l+1)}}\left(\vec{r} \times \vec{\nabla} Y_l^0\right) = \sqrt{\frac{4\pi}{l(l+1)}} \frac{dY_l^0}{d\theta} \hat{e}_{\phi}
\label{Eq12}
\end{eqnarray} 
which are orthonormal with respect to the following inner product:
\begin{eqnarray}
\langle \vec{\Psi}_{\rm{H}, \textit{l}}~\vert ~\vec{\Psi}_{\rm{H}, \textit{l'}}  \rangle = \int_0^{\pi} \vec{\Psi}_{\rm{H}, \textit{l}} \cdotp \vec{\Psi}_{\rm{H}, \textit{l'}} \sin{\theta} d\theta = \delta_{ll'}
\label{Eq13}
\end{eqnarray} 
One tempting choice for the radial function is to use the radial eigenmodes of Eq. (\ref{Eq10}), $\bar{\xi}_{\rm{R}, \textit{n}}$, (where $n$ is the number of radial nodes) as basis functions, i.e. $\Psi_{\rm{R}, \textit{n}} = \bar{\xi}_{\rm{R}, \textit{n}}$. It turns out however, that the expansion of the elasto-magnetic displacement field [see Eq. (\ref{Eq2})] into elastic eigenfunctions is very inefficient. We found that better convergence is realized with 
\begin{eqnarray}
\Psi_{\rm{R}, \textit{n}} (r) &=& r\sqrt{\frac{2}{r_1-r_0}} \cos{\left(\frac{\pi n (r - r_0)}{r_1-r_0}\right)} ~~~\rm{for}~~ \textit{n=}\rm{1,2,...} \nonumber\\
\Psi_{\rm{R}, \textit{n}} (r) &=& r\sqrt{\frac{1}{r_1-r_0}}~~~\rm{for}~~\textit{n=}\rm{0}
\label{Eq14}
\end{eqnarray} 
which obey Eq. (\ref{Eq11}), so that no extra boundary terms in $\vec{L}_{\rm el}$ are needed to preserve the stress-free condition. The basis functions of Eq. (\ref{Eq14}) are orthonormal with respect to the following inner-product:
\begin{eqnarray}
\langle \Psi_{\rm{R}, \textit{n}}~\vert ~\Psi_{\rm{R}, \textit{n'}}  \rangle = \int_{r_0}^{r_1} \Psi_{\rm{R}, \textit{n}} ~\Psi_{\rm{R}, \textit{n'}} \frac{1}{r^2} dr = \delta_{nn'}
\label{Eq15}
\end{eqnarray} 
Combining Eq's (\ref{Eq12}) and (\ref{Eq15}) gives us a series of basis functions that we use in the next section to calculate elasto-magnetic modes of the crust
\begin{eqnarray}
\vec{\Psi}_{ln} (r,\theta) = \Psi_{\rm{R}, \textit{n}}(r) ~\vec{\Psi}_{\rm{H}, \textit{l}}(\theta)
\label{Eq15b}
\end{eqnarray} 
which are orthonormal 
\begin{eqnarray}
\langle \vec{\Psi}_{ln}~\vert ~\vec{\Psi}_{l'n'}  \rangle = \int_{r_0}^{r_1} \int_0^{\pi} \frac{\sin{\theta}}{r^2} \vec{\Psi}_{ln} \cdotp \vec{\Psi}_{l'n'}~ d\theta dr = \delta_{ll'}\delta_{nn'}
\label{Eq15c}
\end{eqnarray} 
Note that the weight function $w$ of Eq. (\ref{Eq3}) takes the form $w(r,\theta) = \sin{\theta}/r^2$.

\subsection{The neutron star model}
We assume that our star is non-rotating and neglect deformations due to magnetic pressure, which are expected to be small. Therefore, we adopt a spherically symmetric background stellar model that is a solution of the Tolman-Oppenheimer-Volkoff equation (TOV equation). We calculate the hydrostatic equillibrium using a SLy equation of state (Douchin \& Haensel, 2001; Haensel \& Potekhin, 2004; Haensel, Potekhin \& Yakovlev, 2007) (see \it{http://www.ioffe.ru/astro/NSG/NSEOS/} \rm for a tabulated version). The model that we use throughout this paper has a mass of $M_*=1.4~M_{\odot}$, a radius $R_*=1.16\cdotp 10^6$ cm, a crust thickness $\Delta R = 7.9 \cdotp 10^4$ cm, a central density $\rho_c = 9.83\cdotp10^{14}~\rm{g~cm^{-3}}$ and cental pressure $P_c = 1.36\cdotp10^{35}~\rm{dyn~cm^{-2}}$. The crustal shear modulus $\mu$ is given by (Strohmayer et al., 1991)
\begin{equation}
\mu = \frac{0.1194}{1+0.595(173/\Gamma)^2} \frac{n(Ze)^2}{a}
\label{16}
\end{equation}
where $n$ is the ion density, $a = (3/4\pi n)^{1/3}$ is the average spa-cing between ions and $\Gamma = (Ze)^2/ak_BT$ is the Coulomb coupling parameter. We evaluate $\mu$ in the limit $\Gamma \rightarrow \infty$.

\begin{figure}[tbp]
  \begin{center}
        \includegraphics[width=3.5in]{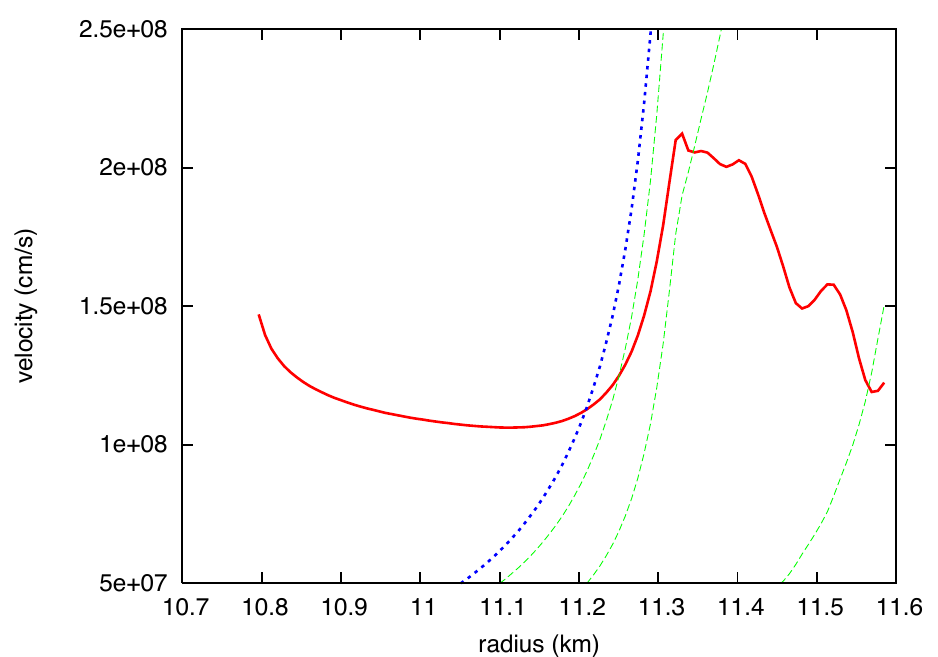}
  \end{center}
  \caption{\small Shear velocity $c_s = \sqrt{\mu / \rho}$ (solid line) versus Alfven velocity $c_A = \sqrt{B^2 / 4\pi \rho}$ for a poloidal field strength of $10^{15}$ G (dotted line). The dashed lines are the radial components of the Alfven velocity, $c_{A, \rm rad} = c_A\cos{\alpha}$, evaluated at (from left to right) $\theta = 69^{\rm{o}}, ~79^{\rm{o}}$ and $89^{\rm{o}}$. Closer to the poles (smaller $\theta$), the field becomes nearly radial and $c_{A, \rm rad} \sim c_A$. The $c_A$-curve shown in this plot is evaluated at the pole ($\theta = 0^{\rm{o}}$), but varies negligibly as a function of $\theta$.}
    \label{Fig0}
\end{figure}

To the spherical star we add a poloidal magnetic field, which we generate as follows: We start with an Euclidean (flat) space into which we place a circular current loop of radius $r_{\rm cl}=0.55~R_*$ and current $I$ and calculate the magnetic field generated by the loop (see e.g. Jackson, 1998). Then we map this field onto the curved space-time of the neutron star, as discussed in section 2. The field is singular near the current loop, however all the field lines which connect to the crust (and thus are physically related to observable oscillations) carry finite field values. This particular field configuration is chosen as an example; there is an infinite number of ways to generate poloidal field configurations. In figure \ref{Fig0} we plot resulting shear- and Alfven velocities in the crust as a function of radial coordinate $r$.

\subsection{Results}
We now use the formalism and equations of the previous sections to calculate elasto-magnetic modes of the magnetar crust. We construct a basis from $N_n$ radial functions $\Psi_{\rm{R} , \textit{n}}(r)$ (see Eq. (\ref{Eq14})) with index $n=0,1,..., N_n-1$, and $N_l$ angular functions $\vec{\Psi}_{\rm{H}, \textit{l}}(\theta)$  (see Eq. (\ref{Eq12})) with even index $l=2,4,..., 2N_l$. These functions provide a set of $N_n \times N_l$ linearly independent basisfunctions $\vec{\Psi}_{ln}$. Using this basis, we solve the matrix equation (\ref{Eq4b}), and reconstruct the normal modes according to Eq. (\ref{Eq2}). \\

\begin{figure}[tbp]
  \begin{center}
        \includegraphics[width=3.5in]{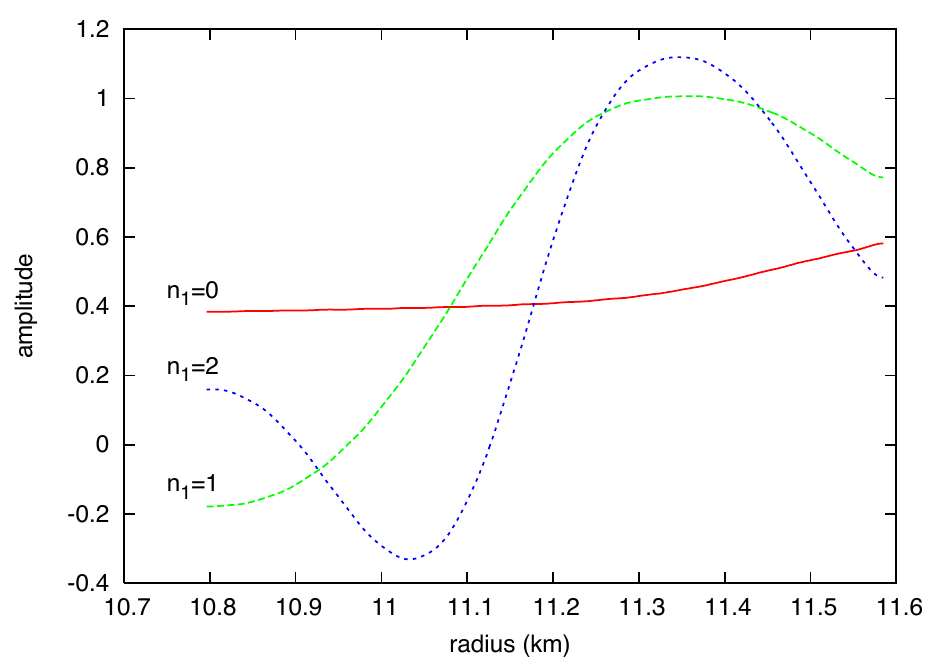}
  \end{center}
  \caption{\small Radial profiles of $l_1=2$ elasto-magnetic modes, evaluated at $\theta = 81^{\rm{o}}$. The vertical scale of individual curves is adapted for visual convenience. }
    \label{Fig1}
\end{figure}

\begin{figure}[tbp]
  \begin{center}
        \includegraphics[width=3.5in]{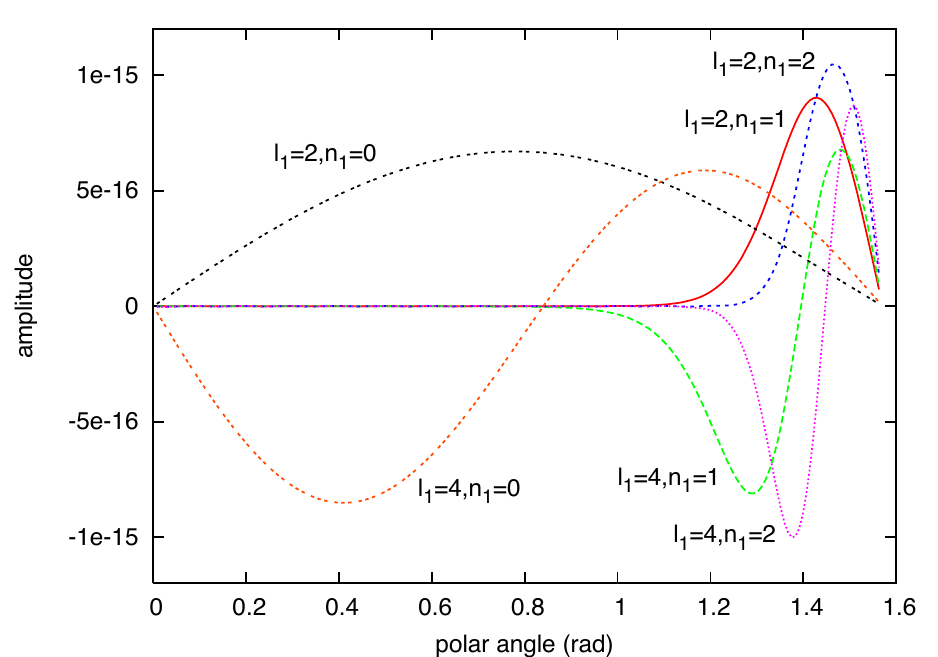}
  \end{center}
  \caption{\small Examples of elasto-magnetic eigenmodes for $B_p=10^{15}$ G (where $B_p$ is field strength at the magnetic pole), as a function of the polar angle $\theta$, evaluated at the crust-core interface. The $n_1=0$ modes are nearly unaffected by the magnetic field and are spread out over the crust, whereas the $n_1>0$ modes are affected strongly by the magnetic field, and are confined to regions near the equator, where the field is horizontal.}
    \label{Fig1b}
\end{figure}

Radial and horizontal cross-sections of a selection of eigenmodes are plotted in figures \ref{Fig1} and \ref{Fig1b}, and table \ref{Table1} contains a list of frequencies. These results are based on a stellar model with a poloidal field strength of $10^{15}$ G at the magnetic pole. For the calculation we used $N_n=35$ radial basis functions and $N_l=35$ angular basis functions. We labeled the modes with integer indices $n_1=0,1,2...$ and $l_1=2,4,6,...$, where $n_1$ is defined as the number of nodes along the $r$-axis and $l_1+1$ is the number of nodes along the $\theta$-axis (including the poles). Note that the index $l_1$, in contrast to $l$, does not signify a spherical harmonic degree since the angular dependence of the elasto-magnetic modes differs from pure spherical harmonics. However, there is a connection between the two indices: the elasto-magnetic mode of degree $l_1$ and order $n_1$, can be interpreted as the magnetically perturbed elastic mode of the same order and (spherical harmonic-) degree. More precisely, if one gradually increases the magnetic field strength, the $n,l$ elastic mode transforms into the elasto-magnetic mode of the same indices, $n_1=n$ and $l_1=l$ (see fig. \ref{Fig4}). It is interesting to note (see fig's \ref{Fig4} and \ref{Fig1b}) that as the field strength increases, modes with $n_1>0$ become more and more confined to a narrow region near the equator (a similar effect was recently observed in the grid-based simulations of Gabler et al. 2011b). In the equatorial regions, the horizontal field creates a magnetic tension-free cavity for modes with radial nodes, which are reflected back towards the equator at higher lattitudes where the field becomes more radial\footnote{A similar  effect is well-known from the study of
waveguides: as the waveguide gets narrower (i.e. as its transverse frequency increases), the propagating
wave may become
evanescent in the longitudinal direction and be reflected}. The $n_1=0$ modes however, having no radial nodes, are virtually insensitive to the magnetic field and are therefore not confined to low lattitudes. 
The field strength-dependence of the eigenfrequencies is illustrated in figure \ref{Fig3}. As we increase the field strength, we find that the increase in frequency $\delta \omega$ for $n_1=0$ modes scales weakly with $B$, i.e. $\delta \omega \propto B^2$. For modes with $n_1>0$, $\delta \omega \propto B^2$ if $B < 5\cdotp 10^{13}$ G, and $\delta \omega \propto B$ if $B > 5\cdotp 10^{13}$ G.\\

\begin{table}[ht]\caption{Normal mode frequencies} 
\centering 
\begin{tabular}{c c c} 
\hline\hline 
mode indices & elastic modes & elasto-magnetic modes \\ 
& ($B=0$ G) & ($B=10^{15}$ G) \\ [0.5ex] 
\hline 
$n_1=0$, $l_1=2$ & 27.42 Hz & 27.61 Hz\\ 
$n_1=0$, $l_1=4$ & 58.16 Hz & 59.14 Hz\\
$n_1=0$, $l_1=6$ & 86.69 Hz & 88.13 Hz\\
$n_1=0$, $l_1=8$ & 114.7 Hz & 116.5 Hz \\
[1ex]$n_1=1$, $l_1=2$ & 895.9 Hz & 954.1 Hz\\
$n_1=1$, $l_1=4$ & 897.4 Hz & 985.7 Hz\\ 
$n_1=1$, $l_1=6$ & 899.7 Hz & 1001.4 Hz\\
$n_1=1$, $l_1=8$ & 902.8 Hz & 1003.4 Hz\\ 
[1ex]
$n_1=2$, $l_1=2$ & 1474.6 Hz & 1607.1 Hz \\ 
$n_1=2$, $l_1=4$ & 1475.7 Hz & 1664.4 Hz\\ 
$n_1=2$, $l_1=6$ & 1477.5 Hz & 1708.1 Hz\\
$n_1=2$, $l_1=8$ & 1479.9 Hz & 1740.4 Hz \\
[1ex] 
\hline 
\end{tabular}
\caption{\small The eigenfrequencies of the non-magnetic crust (second column) versus the eigenfrequencies of the magnetized crust (third column), with a magnetic field of $10^{15}$ G at the polar surface. The elasto-magnetic frequencies were calculated using a basis of $35 \times 35$ basisfunctions $\Psi_{ln}$. }
\label{Table1} 
\end{table}

\begin{figure}[tbp]
  \begin{center}
        \includegraphics[width=3.5in]{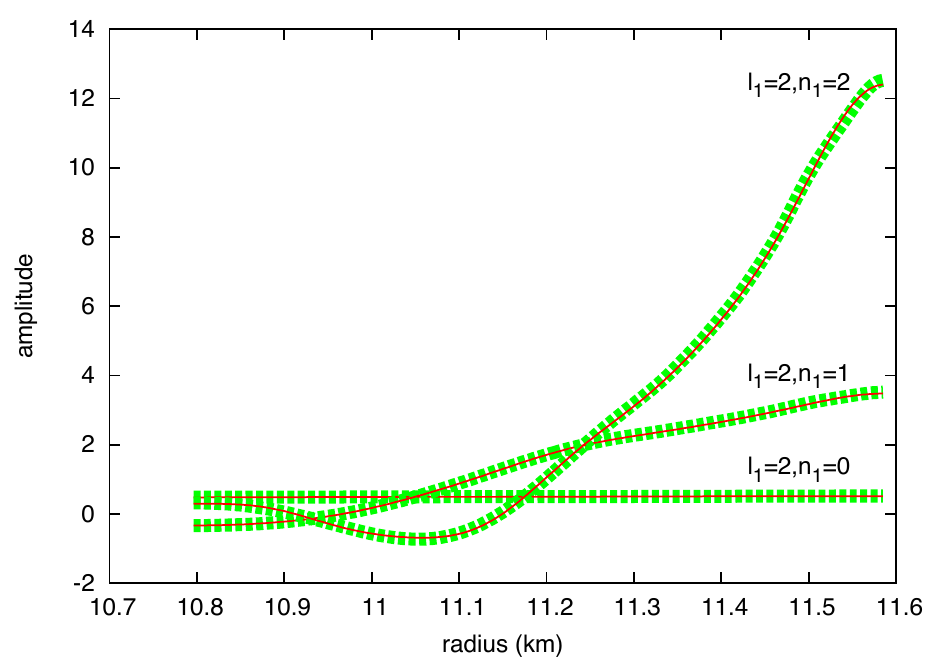}
  \end{center}
  \caption{\small Elastic crustal modes obtained through integration of the elastic equation of motion (thick dashed curves), and the same modes obtained by the series-expansion method (overplotted by the thin solid curve), using $N_n$ radial basis functions. }
    \label{Figel}
\end{figure}

\begin{figure}[tbp]
  \begin{center}
     \includegraphics[width=3.5in]{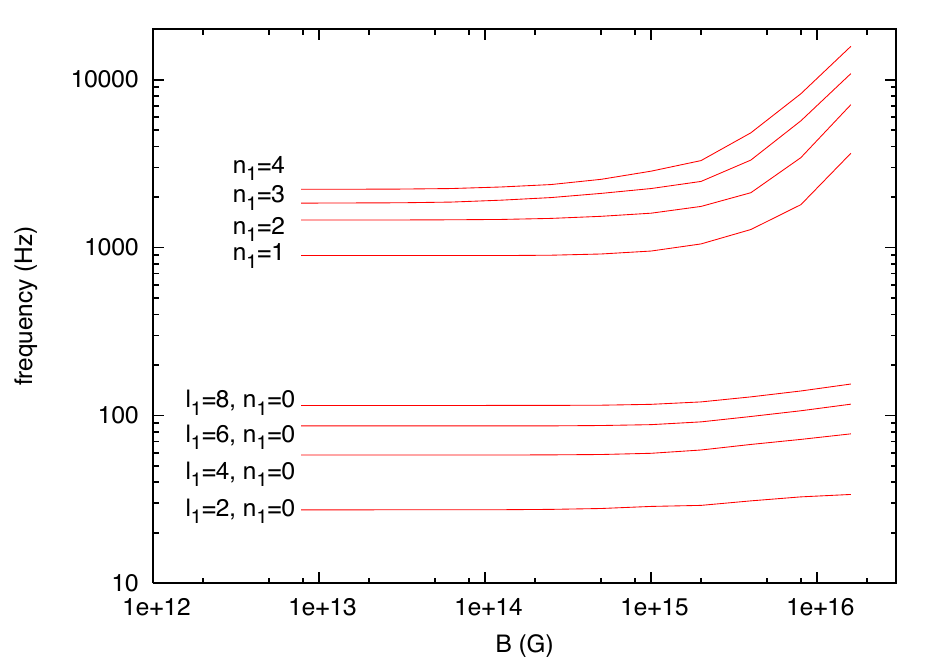}
  \end{center}
  \caption{\small Frequencies as a function of $B$. For $n_1>0$, the frequencies of (low) $l_1$-modes nearly coincide and are therefore collectively indicated with their $n_1$-value, i.e. $n_1=1$, $n_1=2$, etc. Note that high field strengths, the $n_1>0$ frequencies collectively behave as $\omega \propto B$.}
    \label{Fig3}
\end{figure}

\begin{figure}[tbp]
  \begin{center}
     \includegraphics[width=3.5in]{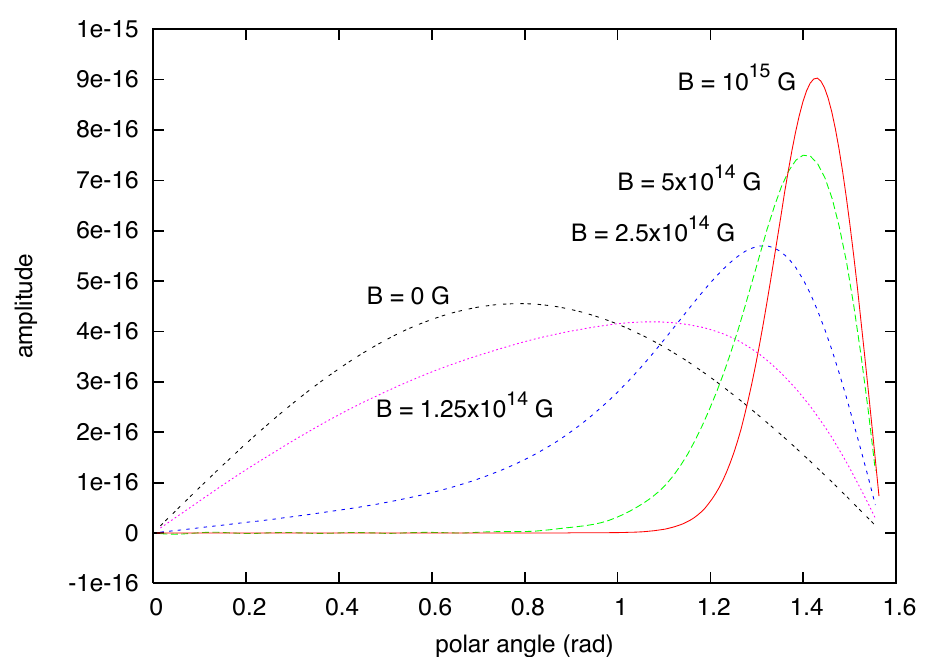}
  \end{center}
  \caption{\small Angular geometry of the $l_1=2$, $n_1=1$ crustal mode (at the crust-core interface), as a function of the magnetic field strength. For zero magnetic field, the curve is identical to the $l=2$ vector spherical harmonic $\Psi_{\rm{H}, \textit{l}}(\theta)$. As the field strength increases, the crustal motion becomes gradually more confined towards the equator.}
    \label{Fig4}
\end{figure}

As a test, we compared the eigenfrequencies and eigenmodes for zero field, $B=0$, to those obtained by a direct integration of the elastic equation of motion of Eq. (\ref{Eq10}).\footnote{The latter works as follows: One starts by assuming harmonic time depence for the displacement $\bar{\xi}$, so that $L_{\rm el} (\vec{\bar{\xi}}) = -\omega^2 \vec{\bar{\xi}}$. Dropping the angular- and time-dependent parts of $\vec{\bar{\xi}}$ on both sides of the equation, one is left with an equation for $\bar{\xi}_{\rm R}$, which is integrated from the bottom of the crust, with corresponding boundary condition, to the surface. This is repeated for different $\omega$ until the surface boundary condition is satisfied; one has found an eigenmode. By repeating this procedure with gradually increasing $\omega$, one obtains a series of eigenmodes and -frequencies.} We find that both frequencies and wavefunctions obtained by the series expansion-method converge rapidly\footnote{Note that in the purely elastic case, $l$ is a good quantum number and the angular basis functions $\vec{\Psi}_{\rm{H}, \textit{l}}(\theta)$ are already solutions to the elastic eigenmode equation. Therefore, for a given $l_1=l$ only the series with the radial basis-functions needs to be considered.} to real values, obtained by integration of Eq. (\ref{Eq10}). E.g. for $N_n = 10$, $n_1=0$ elastic frequencies have a typical error of 0.02\%, while frequencies for modes $n_1<4$ are well within 1\% accuracy. In figure \ref{Figel} we plot elastic eigenfunctions, obtained by both methods. The solutions from the series-expansion method with $N_n = 10$ radial basis functions are nearly indistinguishable from the solutions obtained by direct integration.\\

\begin{figure}[h]
\begin{center}$
\begin{array}{cc}
\includegraphics[width=1.7in]{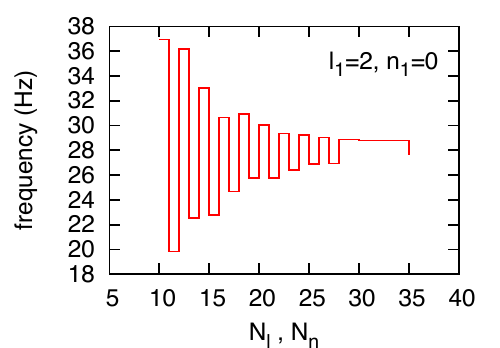} & \includegraphics[width=1.7in]{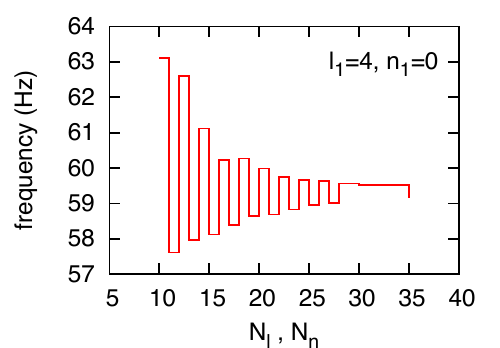} \\
\includegraphics[width=1.7in]{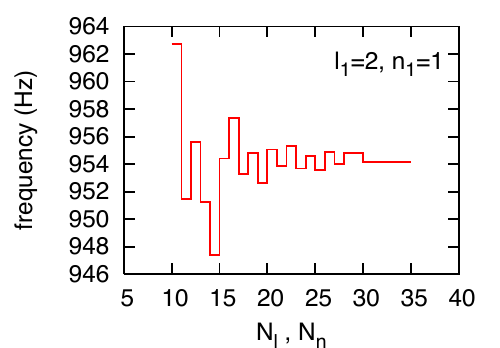} & \includegraphics[width=1.7in]{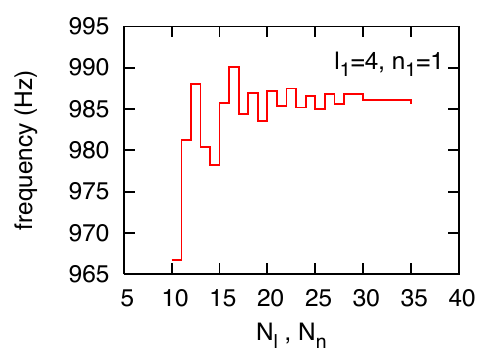}
\end{array}$
\end{center}
\caption{\small Demonstration of convergence for elasto-magnetic frequencies for low-order, low-degree modes as a function of $N_n$ and $N_l$, where we took $N_n = N_l$. The actual number of basisfunctions, $N = N_n \times N_l$, is the square of the value along the x-axis.  }
    \label{Fig2}
\end{figure}

For the full elasto-magnetic equation of motion, Eq. (\ref{Eq1}) with a magnetic field strength of $10^{15}$ G at the pole, we tested the convergence of resulting eigenfrequencies by increasing the number of basis functions $N_n$ and $N_l$ (see figure \ref{Fig2}). We find that, compared to the non-magnetic case, a significant number $N_n$ of radial functions and $N_l$ angular functions is required to get acceptable convergence to stable results. The large number of required radial basis functions can be understood from the fact that the magnetic acceleration $L_B$ (Eq. (\ref{Eq8})) contains delta-functions, arising from the boundary terms. Obviously, one needs many radial basis functions to obtain an acceptable sampling of these boundary terms. The number of computational operations however, is a steep function of the number of basis functions (approximately $\propto (N_l \times N_n)^{3}$), so that computations with large $N_l$ and $N_n$ can become unpractical on ordinary workstations. Although this limits the number of basisfunctions in our calculations, we find that for $N_l, N_n \sim 35$, the scatter in frequencies is typically $\lesssim1\%$ for most modes (figure \ref{Fig2}), and the eigenfunctions $\bar{\xi}_m$ reproduce the orthogonality relation of Eq. (\ref{Eq4g}) with good precision. \\

\section{Core continuum and crust-core coupling}
\subsection{The continuum}

The equation of motion is in this case simply the Alfven wave equation:
\begin{equation}
\frac{\partial^2 \xi(\psi,\chi)}{\partial t^2} = L_{\rm mag}\left[\xi(\psi,\chi)\right],
\label{2.1}
\end{equation}
where $t$ denotes the Schwarzschild time-coordinate. The operator $L_{\rm mag}$ is given in Eq. (\ref{A13}), which we repeat here for convenience
\begin{eqnarray}
L_{\rm mag}\left[\xi ( \psi,\chi)\right] = ~~~~~~~~~~~~~~~~~~~~~~~~~~~~~~~~~~~~~~~~~~~~~~~~~~~~~~~~ \\
\frac{1}{\tilde{\rho}c^2}\sqrt{\frac{g_{tt}}{g_{\chi\chi}}} \frac{B}{4\pi \sqrt{g_{\phi\phi}}} \frac{\partial}{\partial \chi} \left[ \sqrt{\frac{g_{tt}}{g_{\chi\chi}}} g_{\phi\phi} B \frac{\partial}{\partial \chi} \left( \frac{\xi}{\sqrt{g_{\phi\phi}}} \right) \right] \nonumber
\label{2.2}
\end{eqnarray}
Here $g_{tt}$, $g_{\chi \chi}$ and $g_{\phi\phi}$ are the metric terms corresponding to the system of coordinates defined in section 2.

\begin{figure}[htp]
\centering
\includegraphics[width=3.5in]{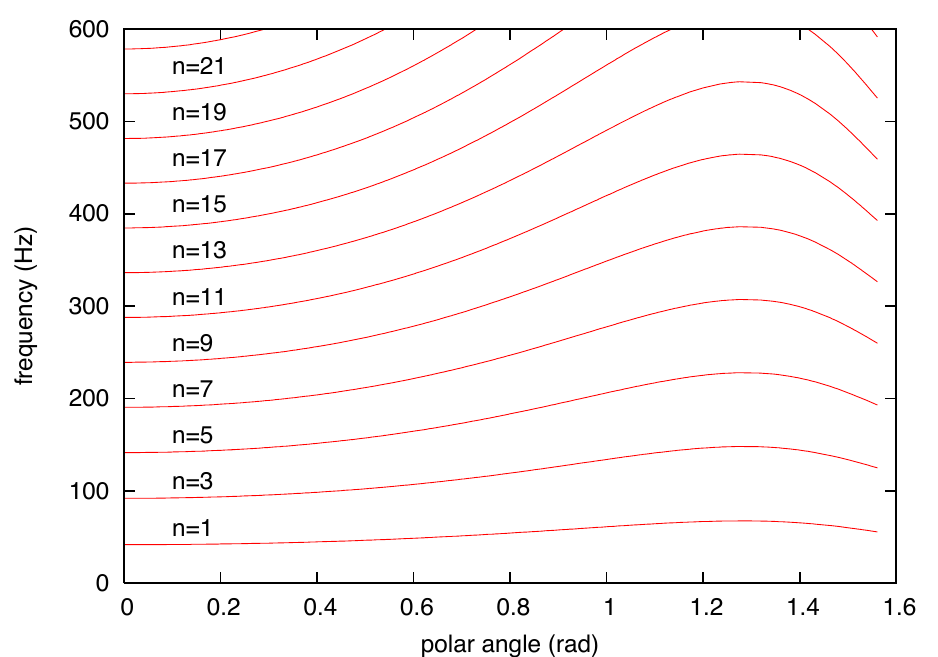}
\caption{\small The curves show the Alfven frequencies $\sigma_n$ as a function of the angle $\theta (\psi)$, the polar angle at which the flux surface $\psi$ intersects the crust. Since we are only considering odd crustal modes, the only Alfven modes that couple to the motion of the star are the ones with an odd harmonic number $n$. This particular continuum was calculated using a poloidal field with a surface value of $B = 10^{15}$ G at the poles.}
\label{core_freqs}
\end{figure}

For determining the spectrum of the core continuum, the appropriate
boundary conditions are $\xi(\chi=\chi_c)=0$, where $\chi_c(\phi)$ marks the
location of the crust-core interface. The full significance
of this boundary condition will become apparent later in this section when we develop the analysis for the
crust-core interaction; see also section 4.2 in vHL11. With this boundary condition, Equation (\ref{2.1}) constitutes a Sturm-Liouville
problem on each separate flux surface $\psi$. 
Using the stellar structure model and magnetic field configuration described in section 3.3, 
we can calculate the eigenfunctions and eigenfrequencies for each flux
surface $\psi$. The reflection symmetry of the stellar model and the magnetic field
with respect to the equatorial plane assures  that the eigenfunctions of
equation~(\ref{2.1}) are either symmetric or anti-symmetric with respect to the
equatorial plane. We can therefore determine the eigenfunctions by integrating
equation~(\ref{2.1}) along the magnetic field lines from the equatorial plane $\chi =
0$ to the crust-core interface $\chi = \chi_c \left( \psi \right)$. Let us consider
the odd modes here for which $\xi \left( 0 \right) = 0$, and solve
equation~(\ref{2.1}) with the boundary condition $\xi \left(\chi_c \right) =
0$ at the crust-core interface; for even modes, the boundary condition is
$d\xi \left( 0 \right)/d\chi=0$. We find the eigenfunctions by means of a
shooting method; using fourth order Runge-Kutta integration we integrate from $\chi
= 0$ to $\chi = \chi_c$. The correct eigenvalues $\sigma_n$ and eigenfunctions
$\xi_n \left( \chi \right)$ are found by changing the value of $\sigma$ until the
boundary condition at $\xi_n$ is satisfied. In this way we gradually increase the
value of $\sigma$ until the desired number of harmonics is obtained. In figure \ref{core_freqs}
we show a typical resulting core-continuum. The continuum is piece-wise, and covers the domains $\sigma = \left[ 41.8, 67.5 \right]$ Hz and $\sigma = \left[ 91.4, \infty \right)$ Hz. Gaps, such as the one between $67.5$ Hz and $91.4$ Hz in fig. \ref{core_freqs}, are a characteristic feature for the type of poloidal field that we employ in this paper, and typically occur at low frequencies (i.e. $\sigma < 150$ Hz). As we discuss in section 4.3, they may give rise to strong low frequency QPOs; see also vHL11 and Colaiuda \& Kokkotas 2011.

According to Sturm-Liouville theory the normalized eigenfunctions $\xi_n$ of
equation~(\ref{2.1}) form an orthonormal basis with respect to the following inner
product:
\begin{eqnarray}
\langle \xi_m, \xi_n \rangle=
\int_0^{\chi_c} r\left( \chi \right) \xi_m\left( \chi \right) \xi_n\left( \chi
\right) d \chi = \delta_{m,n}
\label{eq_inner}
\end{eqnarray}
Where $\delta_{m,n}$ is the Kronecker delta. Noting that the operator $L_{\rm mag} (\xi)$ is in Sturm-Liouville form, one reads off the weight-function $r(\chi)$:
\begin{eqnarray}
r = \sqrt{\frac{g_{\chi \chi}}{g_{tt}}} \frac{ 4 \pi \tilde{\rho}}{B_{\chi}}.
\label{weight}
\end{eqnarray}
 We have checked that the solutions $\xi_n (\chi)$ satisfy the ortho-gonality relations. 

\subsection{Equations of motion for the coupled crust and core}
We are now ready to compute the coupled crust-core motion. In contrast to L07 and vHL11, where the crust was assumed to be an infinitely thin spherical elastic shell, we shall here adopt a crust of finite thickness with realistic structure. We label the lattitudinal location by the flux surface $\psi$ intersecting the crust-core interface, and consider
the crustal axisymmetric displacements $\bar{\xi}_\phi(\psi,r)$, where $r$ is the radial Schwarzschild-coordinate. 
In the MHD approximation, the magnetic stresses enforce a no-slip boundary
condition at the crust-core interface (at $r=r_0$ in the Schwarzschild coordinates of the crust, or $\chi_c$ in the flux-coordinates of the core), such that $\xi \left( \psi, \chi_c
\right) = \bar{\xi} \left( \theta (\psi ),r_0 \right)$ instead of $\xi
\left( \psi, \chi_c \right) = 0$. It is useful to make the following substitution
\begin{eqnarray}
\zeta \left( \psi,\chi \right) \equiv \xi \left( \psi,\chi \right) -
\bar{\xi} \left( \theta(\psi), r_0\right) w\left( \psi,\chi \right)
\label{2.3}
\end{eqnarray}
where we choose the function $w\left( \psi,\chi \right)$ so that (1) it corresponds
to the static displacement in the core and hence satisfies
 $F \left(w\left( \psi,\chi \right) \right) = 0$, and (2)  $w\left( \psi,\chi_c
\right)=1$. From the definition of the operator $F$ it follows that for the odd modes
\begin{eqnarray}
w\left( \psi , \chi \right) =  \sqrt{g_{\phi\phi}}  \int_{0}^{\chi} \sqrt{\frac{g_{\chi \chi}}{g_{tt}}} \frac{K \left( \psi \right)}{g_{\phi\phi} B \left( \psi , \chi' \right)} d\chi'
\label{B2}
\end{eqnarray}
Here the constant $K \left( \psi \right)$ is chosen such that $w\left( \psi,\chi_c
\right) = 1$. The new quantity $\zeta$ from Eq. (\ref{2.3}) now satisfies the boundary condition $\zeta \left(
\psi,\chi_c \right)=0$ and can be expanded into
the Alfven normal modes $\xi_n$ which satisfy the same boundary conditions.

We now proceed by substituting equation~(\ref{2.3}) into equation~(\ref{2.1}) 
thus obtaining a simple equation of motion for $\zeta$
\begin{eqnarray}
\frac{\partial^2 \zeta \left( \psi,\chi \right)}{\partial t^2} - L_{\rm mag} \left( \zeta
\left( \psi,\chi \right) \right) = - w\left( \psi,\chi \right) \frac{\partial^2
\bar{\xi} \left( \theta(\psi),r_0 \right)}{\partial t^2}
\label{2.4}
\end{eqnarray}
We expand $\zeta$ and
$w$ into a series of $\xi_n$'s: 
\begin{eqnarray}
\zeta \left( \psi,\chi,t \right) = \sum_n a_n \left( \psi,t \right) \xi_n \left(
\psi,\chi \right)\\
w \left( \psi,\chi \right) = \sum_n c_n \left( \psi \right) \xi_n \left( \psi,\chi
\right).
\label{2.6}
\end{eqnarray}
Using these expansions, equation~(\ref{2.4}) reduces  to the following equations of motion for the eigenmode
amplitudes $a_n$
\begin{eqnarray}
\frac{\partial^2 a_n \left( \psi \right)}{\partial t^2} + \sigma_n^2 \left( \psi
\right) a_n \left( \psi \right) = - c_n \left( \psi \right) \frac{\partial^2
\bar{\xi}(\psi,r_0)}{\partial t^2}
\label{2.7}
\end{eqnarray}
These equations show how the core Alfven modes are driven by the motion of the
crust. To close the system,
we must address the motion of the crust driven by the hydromagnetic pull from the core:
\begin{eqnarray}
\frac{\partial^2 \bar{\xi}}{\partial t^2} = L_{\rm crust}\left(\bar{\xi}\right) - ~~~~~~~~~~~~~~~~~~~~~~~~~~~~~~~~~~~~~~~~~~~~~\nonumber  \\
\frac{1}{\tilde{\rho}}\left[  \frac{g_{tt}}{g_{\chi \chi}} \frac{\sqrt{g_{\phi\phi}} B^2}{4\pi c^2}\cos{\alpha}\frac{\partial}{\partial \chi} \left( \frac{\xi}{\sqrt{g_{\phi\phi}}} \right) \right] \delta (r - r_{\rm 0}) 
\label{2.8}
\end{eqnarray}
The expression between the square brackets is the hydro-magnetic stress from stellar core acting on the crust, $\alpha$ is the angle between the magnetic field line and the radial coordinate of the star and $L_{\rm crust}\left(\bar{\xi}\right) = L_{\rm mag}\left(\bar{\xi}\right) + L_{\rm el}\left(\bar{\xi}\right)$ is the acceleration of the crustal displacement due to magnetic- and elastic stress (see section 3). We can rewrite this in terms of the coefficients, using Eq. (\ref{2.3}), the definition of $w$, and the expansions and orthogonality relations of Eq's (\ref{Eq4g}) and (\ref{Eq4h}), as:
\begin{eqnarray}
\frac{\partial^2 b_j}{\partial t^2} + \Omega_j^2 b_j = ~~~~~~~~~~~~~~~~~~~~~~~~~~~~~~~~~~~~~\nonumber \\
-\int_{0}^{\pi} \left. \frac{\sqrt{g_{rr}g_{tt}}}{g_{\chi\chi}} \frac{B^2}{2c^2} \cos{\alpha} \left( \sum_n a_n \frac{\partial \xi_n}{\partial \chi} + \right. \right. \\
\left. \left. \sqrt{\frac{g_{\chi\chi}}{g_{tt}}}\frac{K}{B\sqrt{g_{\phi\phi}}}\sum_i b_i \bar{\xi}_i \right) \bar{\xi}_j \right|_{r=r_0} r_0^2 \sin{\theta} d\theta \nonumber
\label{2.9}
\end{eqnarray}
where the coefficients $b_j (t)$ are crustal mode amplitudes defined in Eq's (\ref{Eq4d}) and (\ref{Eq4h}).
Up to this point the derived equations of motion for the crust and the fluid core
are exact. We are now ready to discretize the continuum by converting the integral
of equation~(\ref{2.8}) into a sum over $N$ points $\theta_i$. In order to avoid the
effect of phase coherence (see section 3) which caused drifts in the results
of L07, we sample the continuum randomly over the $\theta$-interval $\left[ 0,
\pi/2 \right]$. In the following, functional dependence of the coordinate $\psi$ or
$\theta \left( \psi \right)$ is substituted by the discrete index $i$ which denotes
the $i$-th flux surface. 

\begin{eqnarray}
\frac{\partial^2 b_j}{\partial t^2} + \Omega_j^2 b_j = ~~~~~~~~~~~~~~~~~~~~~~~~~~~~~~~~~~~~~~~~~~~~~~~~\nonumber \\
-\sum_i \left. \frac{\sqrt{g_{rr, i}g_{tt,i}}}{g_{\chi\chi, i}} \frac{B_i^2}{2c^2} \cos{\alpha_i} \left( \sum_{n,i} a_{n,i} \frac{\partial \xi_{n,i}}{\partial \chi} + \right. \right. \\
\left. \left. \sqrt{\frac{g_{\chi\chi, i}}{g_{tt, i}}}\frac{K_i}{B_i\sqrt{g_{\phi\phi, i}}}\sum_m b_m \bar{\xi}_{m,i} \right) \bar{\xi}_{j,i} \right|_{r=r_0} r_0^2 \sin{\theta_i} \Delta\theta_i \nonumber
\label{2.17}
\end{eqnarray}\\

\begin{eqnarray}
\frac{\partial^2 a_{nk}}{\partial t^2} + \sigma_{nk}^2 a_{nk} = - c_{nk} \sum_j
\frac{\partial^2 b_j}{\partial t^2} \bar{\xi}_{j,k}
\label{2.18}
\end{eqnarray}

These are the equations that fully describe dynamics of our magnetar model. As with
the toy model from section 2 we integrate them using a second order leap-frog
scheme which conserves the total energy to high precision. As a test we keep track
of the total energy of the system during the simulations. Further we have checked
our results by integrating equations (53) and (\ref{2.18}) with the
fourth-order Runge-Kutta scheme for several runs and found good agreement with the leap-frog integration.

\subsection{Results}
Based on the results of vHL11, we expect the following dynamical characteristics to occur; 1) Crustal modes with frequencies that are inside the continuum should undergo resonant absorption, i.e. if such a mode couples efficiently to continuum Alfven modes of the core with similar frequencies, its motion will be damped on rather short time-scales. In the appendix we analytically investigate the efficiency of this coupling and the resulting damping time scales. 2) Late-time behavior of the system will show oscillations near the edges of the continuum; the edge modes. 3) Gaps, as present in the continuum of fig. \ref{core_freqs} will give rise two types of QPOs. First, crustal modes which are inside these gaps will remain undamped, although slightly shifted in frequency due to the interaction with the continuum. Second, edge modes near the edges of the gaps may occur.\\
All of these characteristics were observed in simulations of vHL11, and we expect them to occur in this work. 

\begin{figure}[htp]
\centering
\includegraphics[width=3.5in]{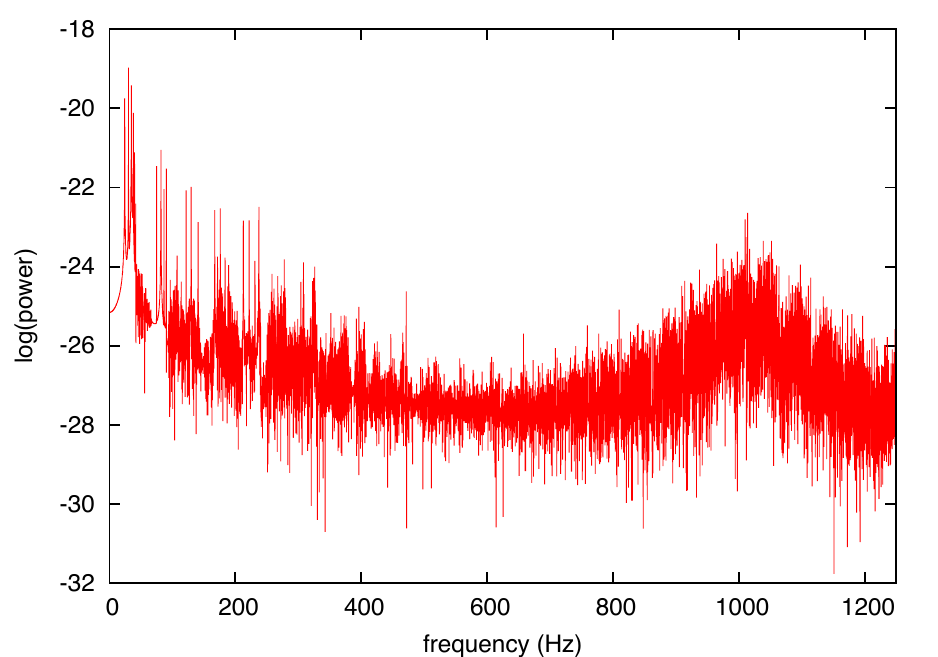}
\caption{\small Power spectrum of the crustal motion.}
\label{fig: structure}
\end{figure}

We consider 16 crustal modes, i.e. $(n,l) = (0,2)$, $(0,4)$, $(0,6)$, $(0,8)$, $(0,10)$, $(0,12)$, $(0,14)$, $(0,16)$, $(0,18)$, $(0,20)$, $(1,2)$, $(1,4)$, $(1,6)$, $(1,8)$, $(1,10)$ and $(1,12)$. We couple these crustal modes to 9000 continuum oscillators, i.e. 300 different flux surfaces, each with 30 Alfven overtones. We start the simulation by initializing the crustal mode amplitudes $b_j = 1$ for all crustal modes, while keeping the continuum oscillators relaxed ($a_{ni}=0$). We evolve the system for 52s in time.\\
In figure \ref{fig: structure} we show the power spectrum which was calculated using the data of the last 26s of the simulation.


\begin{figure}[htp]
\centering
\includegraphics[width=3.5in]{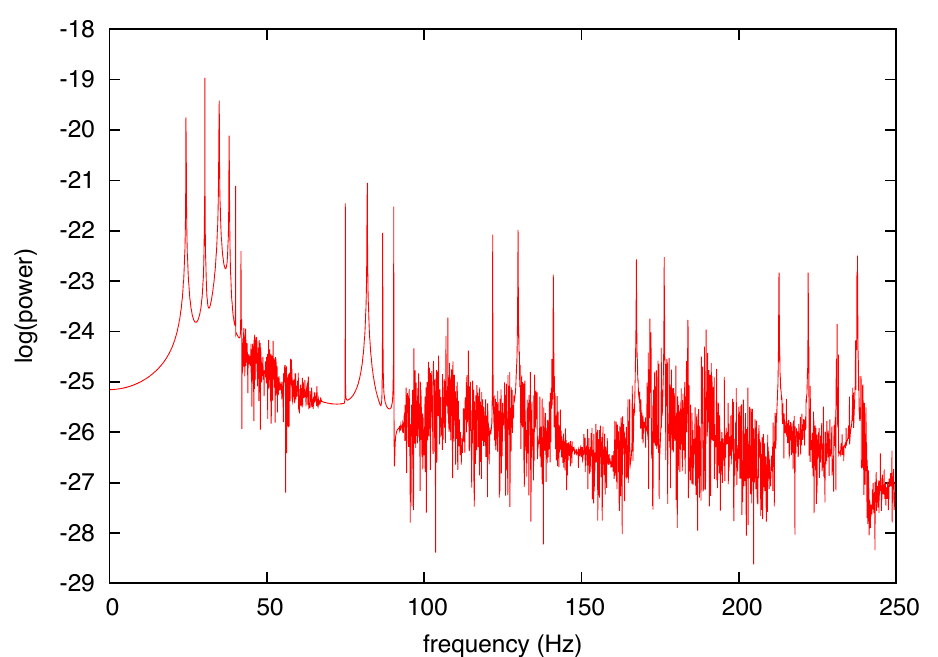}
\caption{\small The same power spectrum, close up.}
\label{spec}
\end{figure}

\begin{figure}[htp]
\centering
\includegraphics[width=3.5in]{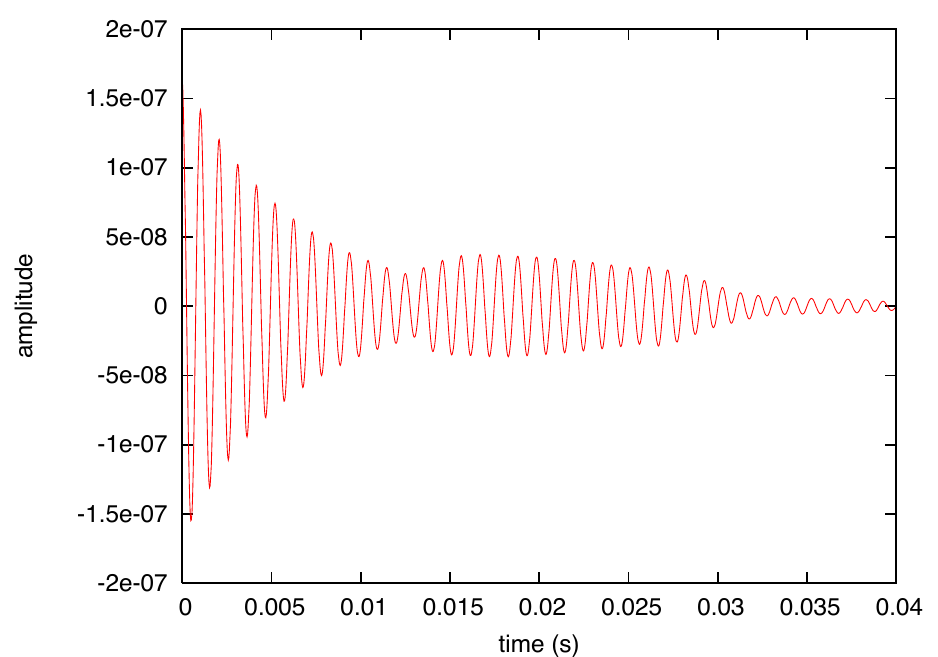}
\caption{\small Displacement of the $l_1=2$, $n_1=1$ mode. The theoretically calculated damping time is $\tau_d = 5.8 \cdotp 10^{-3}$ s. Note the transient increase in the mode amplitude. This is due to the initial Alfven wave train, which is reflected at the equator. }
\label{displ}
\end{figure}

\section{Discussion}
In this paper we have laid out the spectral formalism for computation of general-relativistic torsional
 magnetar oscillations. This method is efficient; a typical simulation of 50 seconds of
the magnetar dynamics (i.e., up to tens of thousands of the oscillatory periods) takes only a few hours
or an ordinary workstation. By contrast, published calculations by groups using alternative finite-different
schemes (Gabler et al. 2011a, Colaiuda \& Kokkotas 2011, Gabler et al. 2011b) use massive computational
resources but still can track at most several tens of the oscillations periods. The second-order symplectic
leap-frog scheme ensures that the energy of the system is conserved with very high accuracy.
Our simulations allow us to investigate which of the oscillatory behavior is long-lived enough ($\sim 100$s) to be
relevant to the observations of QPOS in the tails of giant SGR flares 
(Israel et al. 2005, Strohmayer \& Watts 2006).

One of the puzzling features of the observations are several high-frequency QPOs above $600$ Hz (Watts \& Strohmayer 2006). The thin-crust models of vH11 had strongly suggested that crustal modes of such high frequency should be subject to the strong resonant absorption in the core, even if the core's Alfven modes do not form a mathematical continuum\footnote{This is because the frequencies of  even discrete Alfven modes form a grid, whose characteristic spacing is much less than $600$Hz. At such high frequencies, the grid acts dynamically as a continuum. See vH11 for a more detailed discussion}. In accordance with recent results of Gabler et al. (2011b), we found that some crustal modes are confined to the regions in the crust where the magnetic field is nearly horizontal. Because of this, the coupling to the Alfven modes in the core is reduced relative to the coupling strength estimated in vHL11, however, the coupling is still large enough for the mode energy to be drained on a time-scale small compared to the observed QPOs ($\tau_d \ll 100$ s). Thus the problem of the high frequency QPOs ($>600$ Hz) still stands.\\

\section{Acknowledgements}
This research was supported, in part, by the Leiden Observatory and the Lorentz Institute through internal grants. MvH thanks Monash School of Physics, where part of this research was completed, for hospitality during his extensive visit.

\section*{References}
\begin{footnotesize} \noindent
Barat, C. et al., 1983, A\& A, 126, 400\\
Cerd\'{a}-Dur\'{a}n, P., Stergioulas, N. \& Font, J.~2009, MNRAS, 397, 1607\\
Colaiuda, A., Beyer, H. \& Kokkotas, K.D.~2009, MNRAS, 396, 1441\\
Colaiuda, A. \& Kokkotas, K. D., 2011, MNRAS, 414, 3014C\\
Douchin, F. \& Haensel, P., 2001, A\&A, 380, 151D\\
Duncan, R. C. 1998, ApJ, 498, L45\\
Gabler, M., Cerd\'{a} Dur\'{a}n, P., Font, J. A., M\"{u}ller, E., Stergioulas, N., 2011a, MNRAS, 410L, 37G\\
Gabler, M., Cerd\'{a} Dur\'{a}n, P., Stergioulas, N., Font, J. A., M\"{u}ller, E., 2011b, arXiv: 1109.6233 [astro-ph]\\
Glampedakis K., Samuelsson L., Andersson N., 2006, MNRAS, 371, L74\\
Goedbloed, J.~P.~H. \& Poedts, S.~2004, \textit{Principles of Magnetohydrodynamics}, (Cambridge University Press)\\
Goldreich, P. \& Reisenegger, A., 1992, ApJ, 395, 250G\\
Gruzinov, A., 2008, arXiv: 0812.1570 [astro-ph]\\
Haensel P., \& Potekhin A.~Y., 2004, A\&A, 428, 191H\\
Haensel P., Potekhin A.~Y., Yakovlev D.~G., 2007, \textit{Neutron Stars 1: Equation of State and Structure}, (Springer, New York)\\
van Hoven, M.B. \& Levin, Y., 2011, MNRAS, 410, 1036V (vHL11 in the text)\\
http://www.ioffe.ru/astro/NSG/NSEOS/\\
Israel, G.~L. et al., 2005, ApJ, 628, L53\\
Jackson, J. D., 1998, \textit{Classical Electrodynamics}, 3rd edition, (Wiley)\\
Karlovini, M. \& Samuelsson, L., 2007, CQGra, 24, 3171K\\
Kouveliotou, C., et al., 1999, ApJ, 510L, 115K\\
Landau, L.~D. \& Lifshitz, E.~M.~1976, \textit{Mechanics}, (Pergamon press)\\
Lee, U.~2008, MNRAS, 385, 2069\\
Levin Y., 2006, MNRAS Letters, 368, 35 (L06 in the text)\\
Levin Y., 2007, MNRAS, 377, 159 (L07 in the text)\\
Mastrano, A., Melatos, A., Reisenegger, A., Akg\"{u}n, T., 2011, MNRAS, tmp, 1462M\\
Misner, C. W., Thorne, K. S., Wheeler, J. A., 1973, \textit{Gravitation}, (W.H. Freeman \& Co., San Francisco)\\
Piro, A.~L., 2005, ApJ, 634L, 153P\\
Poedts, S., Hermans, D. \& Goossens, M.~1985, A\&A, 151, 16\\
Samuelsson, L. \& Andersson, N., 2007, MNRAS, 374, 256S\\
Schumaker, B. L. \& Thorne, K. S., 1983, MNRAS, 203, 457S\\
Sotani, H., Kokkotas, K.~D., Stergioulas, N.~2008, MNRAS Letters, 385, 5\\
Steiner W. \& Watts A.~L., 2009, Phys.~Rev.~Letters, 103r1101S\\
Strohmayer, T.~E. \& Watts, A.~L., 2005, ApJ, 632, L111\\
Watts, A.~L. \& Reddy, S.~2007, MNRAS Letters, 379, 63\\
Watts, A.~L. \& Strohmayer T.~E., 2006, ApJ, 637, L117\\
\end{footnotesize}

\appendix
\section{Damped modes}
Now we explore the phenomenon of resonant absorption which occurs in a system where a harmonic oscillator is coupled to a continuum of oscillators. Our aim is to find an analytic estimate for the rate at which the energy of such an oscillator is transferred to the continuum. The objective of this section and the method that we follow, are analogous to a derivation of the quantum mechanical Fermi's Golden Rule, which gives the transition rate from one quantum mechanical eigenstate into a continuum of states. \\
Consider the coupled crust-core dynamics of section 4. The forced motion of the core Alfven modes due to the acceleration of the crust, is 
\begin{eqnarray}
\ddot{a}_n(\psi) + \sigma_n^2(\psi) a_n(\psi) =  - c_n (\psi) ~\ddot{\bar{\xi}}(\psi, r_0) 
\label{Dm1}
\end{eqnarray}
where $a_n(\psi)$ is the displacement of the $n$-th core Alfven harmonic on the flux-surface $\psi$ with frequency $\sigma_n$, $\ddot{\bar{\xi}} (\psi, r_0)$ is the acceleration of the crust at the location where the flux surface $\psi$ intersects the crust, and $c_n(\psi) = \langle w(\psi , \chi) , \xi_n \rangle$ is a coupling constant (see Eq. (\ref{2.6})). Suppose that we keep the system initially fixed in a position where the crust is displaced with amplitude $b_{m,0}$ according to the $m$-th eigenmode, i.e. $\bar{\xi} = b_{m,0} \bar{\xi}_m$, and the continuum oscillators are relaxed; $a_n (\psi) = 0$. At time $t=0$ we release the crust which starts oscillating at frequency $\Omega_m$. Suppose that the damping timescale $\tau_{d,m}$ of the crustal mode is much larger than its period $\tau_m = 2\pi/\Omega_m$, then the crust oscillates at roughly constant amplitude, i.e. $b_m (t) \approx b_{m,0} \cos{\Omega_m t}$. This motion forces the Alfven oscillators according to
\begin{eqnarray}
\ddot{a}_n(\psi) + \sigma_n^2(\psi) a_n(\psi) =  c_n (\psi) ~\Omega_m^2 b_{m,0}~ \bar{\xi}_m (\psi, r_0) \cos{\Omega_m t} 
\label{Dm}
\end{eqnarray}
One can solve the time-evolution of the oscillator $a_n(t)$ using standard techniques (see e.g. Landau \& Lifshitz, Mechanics $\S$22). After a time $t$ the energy per flux surface $\mathcal{E}_n(\psi) = 1/2 (\dot{a}_n^2 + \sigma_n^2 a_n^2)$ absorbed by the oscillator is
\begin{eqnarray}
\mathcal{E}_n(\psi ,t) &=& \frac{1}{2} c_n^2 (\psi)~\Omega_m^4 b_{m,0}^2~ \bar{\xi}_m^2 (\psi, r_0) \left| \int_0^t  \cos{\Omega_m t'} e^{-i\sigma_n t'} dt'     \right|^2
\label{Dm2}
\end{eqnarray}
It is easy to verify that at late times the term between the vertical brackets in Eq. (\ref{Dm2}) becomes narrowly peaked around $\sigma_n = \Omega_m$, so that the bulk of energy is transported to oscillators which are in (near) resonance with the crust. The average rate of energy (per flux surface) transfer $\langle \dot{\mathcal{E}}_n(\psi, t) \rangle$ from the crust to the to the flux surface $\psi$ at time $t$ is $\mathcal{E}_n(\psi ,t)/t$. For sufficiently large $t$ one finds
\begin{eqnarray}
\langle \dot{\mathcal{E}}_n (\psi, t) \rangle \approx \frac{\pi}{4} c_n^2 (\psi)~\Omega_m^4 b_{m,0}^2~ \bar{\xi}_m^2 (\psi, r_0)   \delta (\Omega_m - \sigma_n)
\label{Dm3}
\end{eqnarray}
where $\delta (\Omega_m - \sigma_n)$ is a Dirac delta function. This expression is exact in the limit of $t \rightarrow \infty$. The total rate of energy transfer $\dot{E}$ from the crust to the Alfven continuum is then obtained simply by integrating Eq. (\ref{Dm3}) over $\psi$ and summing over all $n$
\begin{eqnarray}
\dot{E} = \sum_n \int_{\psi_{\rm min}}^{\psi_{\rm max}} \langle \dot{\mathcal{E}}_n (\psi) \rangle d\psi = \sum_{n,k} \frac{\pi}{4}  c_n^2 (\psi_k)~\Omega_m^4 b_{m,0}^2~ \bar{\xi}_m^2 (\psi_k, r_0)    \left. \frac{d\psi}{d\sigma_n}\right|_{\psi = \psi_k}
\label{Dm4}
\end{eqnarray}
here $\psi_k$ denotes flux surfaces that are in resonance with the crustal motion, $\sigma_n(\psi_k) = \Omega_m$. Since for a given $n$, the crustal mode may be in resonance with Alfven modes in several flux surfaces $\psi_k$, the total energy transfer is obtained by summing over the index $k$. Eq. (\ref{Dm4}), which is the analog of the quantum physics' Fermi's Golden Rule, leads to a simple expression for the energy damping timescale $\tau_{E,m}$ ($= 1/2~ \tau_{d,m}$) of the crustal mode
\begin{eqnarray}
\tau_{E,m} \sim \frac{E (t=0)}{\dot{E}} = \left[ \sum_{n,k} \frac{\pi}{2} \Omega_m^2 c_n^2 (\psi_k) \bar{\xi}_m^2 \left. \frac{d \psi}{d \sigma_n}\right|_{\psi = \psi_k} \right]^{-1}
\label{Dm5}
\end{eqnarray}
where $E (t=0) = 1/2 \Omega_m^2 b_{m,0}^2$ is the initial energy of the $m$-th crustal mode. Using numerical simulations, we verified the correctness of Eq. (\ref{Dm5}). Even for very short damping times, i.e. $\tau_d = 2\tau_E \sim 2\pi/\omega_0$, Eq. (\ref{Dm5}) proves remarkably accurate.\\

\end{document}